\documentclass[10pt,compsoc]{IEEEtran}

\ifCLASSOPTIONpeerreview
  \usepackage[nocompress]{cite}
\else
  \usepackage{cite}
\fi
\ifCLASSINFOpdf
\else
\fi
\usepackage{comment}
\usepackage{ragged2e}
\usepackage[nopatch=eqnum]{microtype}                 
\PassOptionsToPackage{warn}{textcomp}  
\usepackage{textcomp}                  
\usepackage{mathptmx}                  
\usepackage{times}                     
\usepackage{cite}                      
\usepackage{tabu}                      
\usepackage{booktabs}                  
\usepackage[table,usenames,dvipsnames]{xcolor}
\usepackage{pgfplots}
\pgfplotsset{compat=1.16}
\usepackage{url}
\usepackage[breaklinks]{hyperref}

\newcommand{\RN}[1]{\uppercase\expandafter{\romannumeral#1}}
\usepackage[T1]{fontenc}
\usepackage{etoolbox}

\usepackage{multirow}
\usepackage{comment}
\usepackage{tabularx}
\usepackage{multicol, blindtext}
\usepackage{enumitem}

\usepackage{caption}



\makeatletter
\renewcommand\@makefntext[1]%
    {\noindent\makebox[0pt][r]{\textsuperscript{\@thefnmark}\,}#1}
\makeatother

\makeatletter
\newcommand*\mysize{%
  \@setfontsize\mysize{8}{9.0}%
}
\makeatother

\begin{document}

\title{\huge{``Being Simple on Complex Issues'' -- Accounts on Visual \\ Data Communication about Climate Change}}

%
%
%
%

\author{Regina Schuster,
        Kathleen Gregory,
        Torsten Möller,
        ~\IEEEmembership{Senior Member,~IEEE},
        and Laura Koesten

}

\ifCLASSOPTIONpeerreview
\markboth{Manuscript submitted to IEEE Transactions on Visualization and Computer Graphics}%
 \fi
%



\IEEEtitleabstractindextext{%
\begin{justify}
\begin{abstract}
Data visualizations play a critical role in both communicating scientific evidence about climate change and in stimulating engagement and action. To investigate how visualizations can be better utilized to communicate the complexities of climate change to different audiences, we conducted interviews with 17 experts in the fields of climate change, data visualization, and science communication, as well as with 12 laypersons. Besides questions about climate change communication and various aspects of data visualizations, we also asked participants to share what they think is the main takeaway message for two exemplary climate change data visualizations. Through a thematic analysis, we observe differences regarding the included contents, the length and abstraction of messages, and the sensemaking process between and among the participant groups. On average, experts formulated shorter and more abstract messages, often referring to higher-level conclusions rather than specific details. We use our findings to reflect on design decisions for creating more effective visualizations, particularly in news media sources geared toward lay audiences. We hereby discuss the adaption of contents according to the needs of the audience, the trade-off between simplification and accuracy, as well as techniques to make a visualization attractive.
\end{abstract}
\end{justify}

\vspace*{-\baselineskip}
}

\maketitle

\IEEEdisplaynontitleabstractindextext

\ifCLASSOPTIONpeerreview
 \fi
%
\IEEEpeerreviewmaketitle

\ifCLASSOPTIONcompsoc
\IEEEraisesectionheading{\section{Motivation}\label{sec:introduction}}
\else
\section{Introduction}
\label{sec:introduction}
\fi

\IEEEPARstart{P}{ublic} understanding of climate change is a persistent and urgent topic of concern. Despite overwhelming scientific consensus regarding the threat of climate change~\cite{bayes_research_2023}, public opinion continues to be divided~\cite{pew_research_center_response_2021} with alarming prognoses coexisting with a lack of public engagement. Potential reasons for this division are as complex as the topic itself, ranging from a lack of emotional interest among the public~\cite{brosch_affect_2021} to trade-offs between short- and long-term benefits~\cite{markman_why_2018}, to a lack of understanding about scientific evidence~\cite{joslyn_explaining_2021}. The `wickedness' of the problem~\cite{head2008wicked} further complicates public understanding, as do the vast amounts of heterogeneous data~\cite{johansson_evaluating_2010}, which are used to create and test climate models, often with a high degree of uncertainty attached. The complexity of climate change, coupled with the mixture of data and methods, makes communication, particularly to lay audiences, a challenge.
Visual communication formats are increasingly used to address these challenges~\cite{oneill_climate_2014}. Climate-related data are visualized in a variety of (at times contentious) ways~\cite{mann2012hockey}, for decision-makers as well as for the public~\cite{schneider_climate_2012}. Despite this, little research explores how laypeople interpret data visualizations related to climate change, particularly those in news sources. One of our goals is to contribute to the inclusivity of data visualizations by studying different stakeholders.
In this study, we investigate how experts and laypeople make sense of climate change data visualizations to improve how data visualizations are utilized to communicate the complexities of climate change to lay audiences.

To this end, we conducted semi-structured interviews with $17$ participants with expertise in either climate change, data visualization, or science communication, as well as with $12$ laypersons. We asked participants to interact with and determine the main takeaway message for two visualizations published in news sources, which had been adapted from well-known graphics in a recent report by the Intergovernmental Panel on Climate Change (IPCC)~\cite{ipcc_summary_2021}. We then showed participants the original IPCC graphics and asked them to make comparisons. Through a thematic analysis of the in-depth interviews, we identify common challenges in making sense of visualizations and potential avenues for creating understandable climate change visualizations in news sources. This work contributes to existing research by providing:
\begin{itemize}[leftmargin=6mm]
    \item insights about how laypeople understand climate change data visualizations and how their takeaway messages differ from messages formulated by experts
    \item expert opinions about the role of news sources and data visualizations in climate change communication
    \item a better understanding of how to cater climate change data visualizations to lay audiences based on (perceived) difficulties or preferences and opinions from experts and laypersons
\end{itemize}

\section{Research Background} 
\label{section:researchbackground}
After a brief introduction to data visualizations for public audiences and visual sensemaking through takeaway messages, we situate our study in research investigating climate change data visualizations from the IPCC and news outlets. 

\subsection{Data Visualizations for Public Audiences}
The visualization community has acknowledged the diversity of the audiences of data visualizations (such as in~\cite{bottinger_reflections_2020,lee_reaching_2020,peck_data_2019}) with an increasing interest in visualizations for the public (for example ~\cite{burns_invisible_2022,park_graphoto_2018,kennedy_feeling_2018,sprague_exploring_2012,pousman_casual_2007}). Besides notions of visualization literacy~\cite{lee_vlat_2017,boy_principled_2014}, a growing body of research concerns itself with the evaluation of data visualizations according to various aspects, including understandability or cognitive learning objectives~\cite{adar_communicative_2021,franconeri_science_2021}, but also affective intents (e.g.,~\cite{lee-robbins_affective_2023, holder_polarizing_2023}) that can shape the readers' opinions, attitudes, and values. 
Despite the increasing research interest in data visualizations for the public (also through qualitative methods), we still know little about how lay audiences engage with data visualizations~\cite{he_enthusiastic_2023}. At the same time, the visualization community has acknowledged the need for more research on climate change data visualization for the public to enhance engagement~\cite{morini_shock_2023}. 
Quispel et al.~\cite{quispel_graph_2016} report a notable study in which they investigated how laypersons and experts in design evaluate graphs according to different aspects. They found that the readers' evaluations of attractiveness appeared to be predicted by perceived ease of use and self-assessed familiarity, with this effect being stronger for laypersons than for experts. 
While we also examine expert vs. lay opinions, our work takes into account various expertise and focuses on how different audiences understand climate change data visualizations and their views concerning visual appeal, understandability, and trustworthiness.

\subsection{Visual Sensemaking and Takeaway Messages}
Many factors influence how humans understand or make sense of data visualizations, including graph formats, visual characteristics, and readers' knowledge about graphs and contents~\cite{lee_investigation_2017}. Peck et al.~\cite{peck_data_2019} studied attitudes and perceptions of data visualizations among participants from rural Pennsylvania and found that educational background, political affiliation, and personal experience influence how participants perceived and interpreted the visualizations. 
Readers' interpretation is also known to be influenced by visualization design choices, which ``have the power to embolden a message or to mute it''~\cite{landers_storytelling_2019}. Previous studies have investigated the effects of various design aspects on sensemaking, insight extraction, and viewer takeaways, for example, the usage of pictographs~\cite{burns_designing_2022} or the visual arrangement of bar charts~\cite{xiong_visual_2021}. Sensemaking has been studied across various fields of research, such as psychology or decision-making (for example,~\cite{malakis_sensemaking_2013,klein_data-frame_2007}), but also human-computer interaction (for example,~\cite{russell_cost_1993}).
However, sensemaking processes and understanding of data visualizations are not easy to measure~\cite{burns_communicative_2022}. The formulation of takeaway messages and with that the assessment of how well readers can 'accurately decode [the] message/insight' of a visualization is discussed as an evaluation method of visualizations in work by Adar and Lee~\cite{adar_communicative_2021} and used as a proxy for measuring readers' understanding of data visualizations (such as in~\cite{bateman_useful_2010} or ~\cite{gammelgaard_ballantyne_images_2016}). Little is known about how different audiences formulate takeaway messages and make sense of climate change data visualizations.

\subsection{Climate Change Data Visualizations by the IPCC}
The IPCC publishes regular assessments summarizing the latest science related to climate change. With $195$ member countries as of $2023$ and thousands of contributing scientists and experts, the IPCC is one of the most established sources of climate knowledge~\cite{ipcc_about_nodate}. Data visualizations have become a crucial part of climate change communication, also within the IPCC: "data visuals are integral to the work of the IPCC''~\cite[p. 23]{corner_principles_2018}. The importance of visuals in IPCC publications has also been studied scientifically. Pidcock et al.~\cite{pidcock_evaluating_2021} conducted an online survey and semi-structured interviews with IPCC authors to investigate how they engage non-expert audiences with climate change topics. Participants mentioned that effective visualizations are a critical factor in public engagement; $35$\% of respondents named IPCC figures and $21$\% scientific data-driven figures as essential tools in public understanding. 
      
The IPCC itself has acknowledged that non-scientists often struggle to interpret their data visualizations correctly~\cite{corner_principles_2018}. However, creating data visualizations that are both scientifically correct and accessible for lay readers is challenging. As Wozniak states, "visual science communication on climate change has always been walking a thin line between scientific accuracy on the one hand and reducing complexity for public understanding and engagement on the other''~\cite[p. 136]{wozniak_stakeholders_2020}. IPCC visualizations fulfill a variety of functions. They depict complex facts and simulated projections and provide guidelines to (non-expert) decision-makers while needing to conform to aesthetic guidelines~\cite{wozniak_stakeholders_2020}. In a study involving both IPCC authors and a group of students, Harold et al.~\cite{harold_communication_2020} found that IPCC authors generally know that lay audiences have difficulty comprehending their visuals.

Some works explicitly focus on identifying how understandable IPCC graphics are for their intended audiences. Fischer et al.~\cite{fischer_how_2018} examined how accurately a sample of visitors at a climate change conference and a sample of students interpreted IPCC visualizations. Both groups showed a low interpretation accuracy; respondents from the conference sample also incorrectly assessed which graphs they had interpreted correctly (meta-cognition). Fischer et al.~\cite{fischer_when_2020} evaluated graph comprehension among decision-makers from $54$ countries using two IPCC graphs, which either employed or violated principles of intuitive design (such as when higher numbers indicate lower climate change impact). Counter-intuitive visualizations were systematically misinterpreted. McMahon et al.~\cite{mcmahon_unseen_2015} further found that study participants representative of the IPCC's target audience (such as interested scholars, governments, and non-governmental intermediaries) had difficulties identifying the types of uncertainty present in IPCC visualizations. 
      
Harold et al.~\cite{harold_enhancing_2017} recommend four pillars that the IPCC should consider to create accessible data visualizations: identification of a message, assessment of the audience's prior knowledge, usage of familiar design formats, and evaluation with the target audience. While current efforts to improve the accessibility of their data visualizations seem to have paid off with informal reactions to a $2021$ report being largely positive~\cite{gaulkin_why_2021}, scientific work has been criticizing IPCC graphics for being inaccessible for non-specialists~\cite{harold_cognitive_2016}.

\subsection{Climate Change Data Visualizations in News Media}
In a longitudinal analysis of climate change imagery, O'Neill~\cite {oneill_more_2020} examined over a thousand images published in UK and US newspapers between $2001$ and $2009$. While the number of visuals increased rapidly after $2005$, scientific data visualizations, such as charts or maps, were still relatively uncommon in the corpus. Research on climate change visuals in news sources tends to focus on i. the content or ii. the public perception. While the former studies the themes or frames (see examples like~\cite{oneill_more_2020,hopke_visualizing_2018,wessler_global_2016}), the latter focuses on the perception of climate change visuals and people's engagement with them (for example~\cite{wang_public_2018,feldman_is_2018,metag_perceptions_2016}). Most of this research addresses the use of imagery in general, focusing on photographs and lacking a clear distinction of data visualizations. Analyses of the design spaces of climate change data visualizations~\cite{windhager_inconvenient_2019, nocke_visualization_2008} do not focus solely on news sources. Little is known about how public audiences understand climate change data visualizations compared to other forms of imagery. At the same time, analyses investigating the understandability of data visuals about climate change often use graphics published by the IPCC rather than visualizations from news sources.

\section{Methodology} 
To investigate how data visualizations can be better utilized to communicate the complexities of climate change to different audiences, we conducted interviews with experts and laypeople. The rationale behind this is outlined in the subsequent sections. 

\subsection{Research Focus and Questions} 
\label{section:researchfocus}
With IPCC materials being used as the basis for the public discourse on climate change~\cite{dudman_ipcc_2021}, many analyses testing the understandability of such visuals commonly use IPCC figures. This leads to the question of whether news media sources frequently publish data visualizations about climate change that differ from the ones by the IPCC. Hence, we investigated $85$ of the most popular online news sources from the UK, the USA, and Germany for articles that were released in response to the Working Group I contribution to the IPCC's Sixth Assessment Report ``Climate Change 2021: The Physical Science Basis'' (AR6-WGI). 
$132$ articles met our selection criteria (described in the supplementary material) and were examined for data visualizations. While the Summary for Policy Makers of the AR6-WGI (SPM-AR6-WGI) includes ten data visualizations, five of those were rarely depicted in news media sources; the other five were more frequently used. Our news article sample contained $47$ data visualizations, from which $34$ were modified/recreated versions of the originals from the AR6-WGI. Even for this sample consisting only of direct reports about an IPCC publication, there were more recreations of IPCC visualizations in news sources than copies. Considering that studies focusing on the understandability of climate change data visualizations from news sources are still scarce, we identify a need for investigations going beyond IPCC material.
This analysis was conducted in November $2021$; details can be found in the supplementary material.

This study aims to better understand experts' and laypeople's perceptions of how data visualizations can be created to meet the needs of different audiences. We used newspaper representations of climate change data visualizations from the IPCC as discussion prompts in interviews with members of groups that commonly create or use them. Research questions RQ-1 and RQ-2 account for the reader's perspective, exploring how different audiences understand climate change visualizations from different sources. Research question RQ-3 was formulated to represent the creator's point of view and investigates how data visualizations can be designed and used in different channels to address the needs of different audiences.

\begin{enumerate}[leftmargin=10mm]
    \item [\textbf{RQ-1:}] How do laypeople and experts make sense of climate change data visualizations, and what do they take away as a main message?
    \item [\textbf{RQ-2:}] What are common challenges people face when interpreting data visualizations about climate change?
    \item [\textbf{RQ-3:}] How can data visualizations about climate change be designed to support readability, understandability, and trustworthiness? 
\end{enumerate}

\subsection{Interview Study Participants} 
Most studies investigating climate change data visualizations derive implications from interviewing or surveying IPCC authors, and typical readers of IPCC reports such as policymakers (for example \cite{pidcock_evaluating_2021, fischer_when_2020, harold_communication_2020}). Contrary to those, we aim to study different perspectives on this issue. Climate change is a complex topic, and creating data visualizations that are accurate and understandable is a problem that can only be tackled by involving different stakeholders. By combining different perspectives and types of expertise, we explore ``different ways of knowing'', how particular views might dissent or coincide, and how they can be added together. 

We opted to interview experts as they are potentially involved and experienced in creating data visualizations and not only have opinions about their design but also have an interest in outreach and creating understandable visuals. Experts were selected according to their affiliation with one of the following fields of expertise: climate change (\textit{climate}), data visualization (\textit{vis}), and science communication (\textit{scicomm}). Contrary to previous studies, we expand our sample to \textit{climate} experts in general instead of solely IPCC authors to also account for opinions from persons who do not have any stake in IPCC publications and might be potentially less biased. We also include \textit{vis} and \textit{scicomm} experts in our study, as \textit{vis} experts bring expertise on what works technically and in terms of perception, and \textit{scicomm} experts have insights on how lay audiences can be reached. We further interviewed a sample of laypeople (\textit{lay}) to compare the expert opinions with their viewpoints. We consider a person to be ``lay'' if they have no professional expertise in climate change, data visualization, or science communication. The study was approved by our institution's ethics committee.

\begin{figure}[b]
\vspace{-3mm}
 \centering 
 \includegraphics[width=\columnwidth]{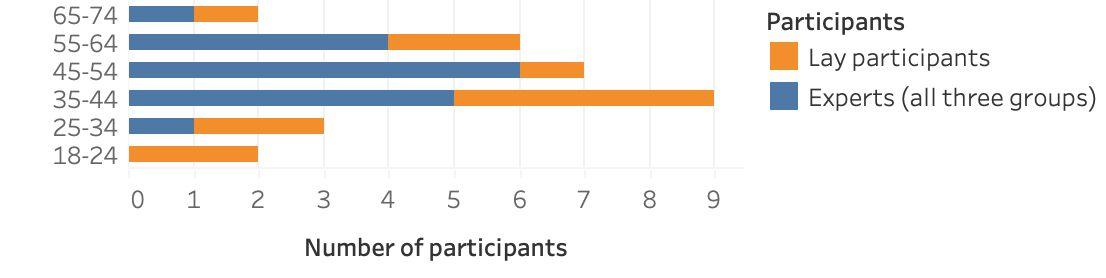}
 \caption{Participants' age ranges}
 \vspace{-1mm}
 \label{fig:age}
\end{figure}

We used purposive sampling within an expert network to contact participants of the expert group. Besides their expertise in either \textit{climate}, \textit{vis}, or \textit{scicomm}, all participants are international experts and known scientists who are experienced in their field and are at an advanced stage in their careers. Experts with a scientific background regularly publish in peer-reviewed journals. Further inclusion criteria included proficiency in English and diversity factors concerning age and gender. \autoref{fig:age} shows participants' age ranges. A total of $17$ experts from Europe and North America were interviewed, from which $6$ had a \textit{climate}, $5$ a \textit{vis}, and $6$ a \textit{scicomm} background. As seen in~\autoref{tab:jobroles}, experts' job roles and research areas or disciplines span several areas. Interviewees were asked about their expertise in the other fields; for example, \textit{climate} experts were asked to rate their experience in \textit{scicomm} and \textit{vis} on a scale from $1$ (very low) to $10$ (very high). \autoref{tab:expertise} shows an overview of the median values within each expert group.

\textit{Lay} participants were recruited using a combination of purposive and convenience sampling, resulting in $12$ interviews. Participants were selected based on diversity factors (age, gender, education, occupation) and contacted via email or phone. We did not define it as a necessity that \textit{lay} interviewees consume specific news sources or data visualizations or have any interest in climate change. \autoref{tab:lay-demo} shows the participants' self-reported age group, job roles, education, and news media usage, including social media, online news sources/apps, television, radio, and print newspapers. All \textit{lay} participants claimed that climate change is something they think about, with reasons spanning the perceived urgency and affectedness, anger towards politics and regulations, and the reporting about activism and nature catastrophes in news sources. Interviewees received no compensation for their participation.

\begin{table}[tb]
\footnotesize
\caption{Experts' self-reported job roles \& research areas}
\centering
\label{tab:jobroles}%
    \begin{tabular}{p{0.5\columnwidth}p{0.40\columnwidth}}
        \toprule
        \multicolumn{1}{c}{\textbf{Job role}} & \multicolumn{1}{c}{\textbf{Research area}} \\
        \midrule
        \multicolumn{2}{c}{Climate change experts (\textit{climate})} \\
        \midrule
        Director of a Climate Research Institute, IPCC author & Climate/ocean modeling \\
        Department Head of Sociology & Social policies \\
        Professor for Climate Science & Physical climate science \\
        Sociologist & Environmental issues \\
        Professor, Faculty Vice Head & Environmental science \\
        Professor in European Ethnology & Urban anthropology \\
        
        \midrule
        \multicolumn{2}{c}{Data visualization experts (\textit{vis})} \\
        \midrule
        Research Scientist & Data visualization, analytics \\
        Retired Visualization Researcher & Data visualization \\
        Professor of Visual Data Science & Data visualization, expl. AI \\
        Professor in Visualization & Visualization \\
        Professor & Visualization \\
        
        \midrule
        \multicolumn{2}{c}{Science communication experts (\textit{scicomm})} \\
        \midrule
        Researcher, Chief Scientific Officer & Modeling and simulation \\
        Head of Communications & Communication \\
        Social Science Method. Professor& Political science \\
        Company Head & Communication \\
        Creative Director & Design, advertising \\
        CEO of Information Design Firm & Information design \\
        \bottomrule
    \end{tabular}
\end{table}

\begin{table}[tb]
\footnotesize
  \caption{Experts' self-reported expertise rated on a scale from $1$ (very low) to $10$ (very high), median values per group}
  \label{tab:expertise}
  \centering
    \begin{tabular}{p{0.5\columnwidth}>{\centering}p{0.1\columnwidth}>{\centering}p{0.1\columnwidth}>{\centering\arraybackslash}p{0.1\columnwidth}}
        \toprule
   & \multicolumn{3}{c}{\textbf{Expertise (median values)}}\\
  Expert group & \textit{climate} & \textit{vis} & \textit{scicomm} \\

  \midrule
  Climate change (\textit{climate}) & $10$ & $6$ & $7$\\
  Data visualization (\textit{vis}) & $3$ & $10$ & $7$\\
  Science communication (\textit{scicomm}) & $4$ & $7$ & $10$ \\
  \bottomrule
  \end{tabular}%
\vspace{-3mm}
\end{table}

\begin{table*}[ht]
\scriptsize
\rowcolors{3}{white}{gray!15}
\caption{\textit{Lay} participants' self-reported age group, job role, education, news usage, opinions regarding the shown visualizations (X: participant uses a source, \textbf{X}: primary source; colors: participant evaluated the visualization as good [green], mediocre [yellow], bad [red])}
\centering
\label{tab:lay-demo}%
\begin{tabular}{p{0.09\columnwidth}p{0.065\columnwidth}p{0.345\columnwidth}p{0.23\columnwidth}p{0.01\columnwidth}p{0.01\columnwidth}p{0.01\columnwidth}p{0.01\columnwidth}p{0.01\columnwidth}p{0.25\columnwidth}p{0.25\columnwidth}}
        \toprule
         & & & & \multicolumn{5}{c}{\textbf{Used news media sources}} & \multicolumn{1}{c}{\textbf{Opinion Vis}} & \multicolumn{1}{c}{\textbf{Opinion Vis}} \\
        \textbf{ID} & \textbf{Age} & \textbf{Job role} & \textbf{Education (highest)} & \multicolumn{1}{c}{Social\footnotemark} & \multicolumn{1}{c}{Online} & \multicolumn{1}{c}{TV} & \multicolumn{1}{c}{Radio} & \multicolumn{1}{c}{Print} & \multicolumn{1}{c}{\textbf{BBC (\autoref{fig:bbc})}} & \multicolumn{1}{c}{\textbf{Guardian  (\autoref{fig:guardian})}} \\
        \midrule   
        LAY-1 & 18-24 & Apprentice as a caregiver & Intermediate school &  
        \multicolumn{1}{c}{X} & \multicolumn{1}{c}{X} & \multicolumn{1}{c}{} & \multicolumn{1}{c}{\textbf{X}} & \multicolumn{1}{c}{} & \cellcolor{red!25}Not understandable & \cellcolor{yellow!25}Interesting, too bold \\

        LAY-2 & 18-24 & Waiting for a place at university & Grammar school &  
        \multicolumn{1}{c}{\textbf{X}} & \multicolumn{1}{c}{} & \multicolumn{1}{c}{} & \multicolumn{1}{c}{X} & \multicolumn{1}{c}{} & \cellcolor{green!25}Clear call to action & \cellcolor{red!25}Informative, too pale \\
        
        LAY-3 & 25-34 & Master student & Bachelor &  
        \multicolumn{1}{c}{\textbf{X}} & \multicolumn{1}{c}{X} & \multicolumn{1}{c}{X} & \multicolumn{1}{c}{X} & \multicolumn{1}{c}{} & \cellcolor{yellow!25}Would not look at it & \cellcolor{red!25}Not understandable \\
        
        LAY-4 & 25-34 & Job seeking & Master &  
        \multicolumn{1}{c}{X} & \multicolumn{1}{c}{\textbf{X}} & \multicolumn{1}{c}{X} & \multicolumn{1}{c}{X} & \multicolumn{1}{c}{} & \cellcolor{yellow!25}Fine, details missing & \cellcolor{red!25}Untrustworthy design \\
        
        LAY-5 & 35-44 & Clerk in accounting & Grammar school &  
        \multicolumn{1}{c}{X} & \multicolumn{1}{c}{X} & \multicolumn{1}{c}{\textbf{X}} & \multicolumn{1}{c}{} & \multicolumn{1}{c}{} & \cellcolor{green!25}Clear, not overloaded & \cellcolor{green!25}Clear, good design \\
        
        LAY-6 & 35-44 & Geriatric nurse, nanny & Intermediate school &  
        \multicolumn{1}{c}{X} & \multicolumn{1}{c}{} & \multicolumn{1}{c}{X} & \multicolumn{1}{c}{\textbf{X}} & \multicolumn{1}{c}{} & \cellcolor{yellow!25}Too scientific & \cellcolor{red!25}Untrustworthy design \\
        
        LAY-7 & 35-44 & Dentist & Doctorate &  
        \multicolumn{1}{c}{} & \multicolumn{1}{c}{\textbf{X}} & \multicolumn{1}{c}{X} & \multicolumn{1}{c}{} & \multicolumn{1}{c}{X} & \cellcolor{red!25}Details missing & \cellcolor{red!25}Untrustworthy design \\
        
        LAY-8 & 35-44 & Goldsmith & Master goldsmith & 
        \multicolumn{1}{c}{\textbf{X}} & \multicolumn{1}{c}{X} & \multicolumn{1}{c}{X} & \multicolumn{1}{c}{X} & \multicolumn{1}{c}{} & \cellcolor{red!25}Not understandable & \cellcolor{red!25}Untrustworthy design \\
        
        LAY-9 & 45-54 & Manager of a biotech firm & Habilitation &  
        \multicolumn{1}{c}{X} & \multicolumn{1}{c}{\textbf{X}} & \multicolumn{1}{c}{} & \multicolumn{1}{c}{} & \multicolumn{1}{c}{} & \cellcolor{yellow!25}Fine, details missing & \cellcolor{yellow!25}Fine, details missing \\
        
        LAY-10 & 55-64 & Manager of a family business & Diploma &  
        \multicolumn{1}{c}{} & \multicolumn{1}{c}{X} & \multicolumn{1}{c}{X} & \multicolumn{1}{c}{} & \multicolumn{1}{c}{\textbf{X}} & \cellcolor{yellow!25}Unclear color choice & \cellcolor{red!25}Untrustworthy design \\
        
        LAY-11 & 55-64 & Retired machine fitter & Secondary school &  
        \multicolumn{1}{c}{} & \multicolumn{1}{c}{} & \multicolumn{1}{c}{\textbf{X}} & \multicolumn{1}{c}{X} & \multicolumn{1}{c}{} & \cellcolor{yellow!25}Would not look at it & \cellcolor{yellow!25}Not interesting \\
        
        LAY-12 & 65-74 & House-wife & Secondary school &  
        \multicolumn{1}{c}{} & \multicolumn{1}{c}{} & \multicolumn{1}{c}{X} & \multicolumn{1}{c}{X} & \multicolumn{1}{c}{\textbf{X}} & \cellcolor{yellow!25}Not easy for everyone & \cellcolor{green!25}Good, message in text \\
        
        \bottomrule
    \end{tabular}
    \vspace{-3mm}
\end{table*}

\subsection{Interview Procedure and Schedule} 
\label{section:interviews}

In semi-structured interviews, participants were asked to discuss climate change data visualizations. Interviews were conducted and recorded online via Zoom or in person according to the participants' preferences. The interviews took $30$-$60$ minutes each and followed a set structure, with some questions being only posed to a specific group: \textbf{(i) Introduction \& consent}. \textbf{(ii) Demographics} \textbf{(iii) Climate change:} \textit{climate} experts were asked about their opinion concerning climate change awareness and understanding among the public and actions taken by companies and governments; \textit{lay} participants were asked about their interest in climate change. \textbf{(iv) Data visualizations:} Participants were asked about aspects of climate change data visualizations and common interpretation difficulties. \textbf{(v) Example visualizations:} Questions included aspects of understandability, improvement potential, and personal preference concerning design choices. Questions were formulated broadly and intended to serve as discussion prompts to identify aspects the interviewees regarded as important. Further questions were included based on the individual discussion focus. Rating questions evaluating aspects on a scale were used to understand the participants' assessments, often leading to more in-depth explanations. The interview schedule can be found in the supplementary material.

Two pairs of data visualizations were used to discuss aspects such as understandability, quality, or trustworthiness. The visuals were selected from the sample of news representations of IPCC graphics (see \autoref{section:researchfocus}). The supplementary material provides an overview of the frequency of copies and recreations of the ten figures in the SPM-AR6-WGI in the investigated online articles. We picked two visualizations that were most often recreated. After discussing the news version, the original IPCC graphic was shown to the participants so they could make comparisons on design choices. If the news version depicted only a part of the original, this was stated, and a zoomed-in version depicting the same data was provided. For \textit{lay} participants, we provided the visualizations in German. \autoref{fig:bbc} (recreation by BBC News at bbc.co.uk/news~\cite{BBC}, referred to as \hyperref[fig:bbc]{Visualization BBC or Vis BBC}) and \autoref{fig:ipcc1} (original by the IPCC~\cite{ipcc_summary_2021}) show the first visualization pair; \autoref{fig:guardian} (recreation by the Guardian~\cite{the_guardian_as_2021}, referred to as \hyperref[fig:guardian]{Visualization Guardian or Vis Guardian}) and \autoref{fig:ipcc2} (original by the IPCC~\cite{ipcc_summary_2021}) show the second visualization pair.

For the news versions of the example visualizations, we asked the participants: ``What do you think is the main takeaway message of this visualization?''. We did not provide explanations or examples of takeaway messages to avoid priming the responses. \hyperref[fig:bbc]{Vis BBC} was shown to all \textit{lay} participants ($12$) and experts ($17$); \hyperref[fig:guardian]{Vis Guardian} was shown to all \textit{lay} participants and $6$ experts for time reasons. Two \textit{lay} participants did not formulate a message for \hyperref[fig:guardian]{Vis Guardian}, stating that they did not understand the graph. This results in $45$ messages for thematic analysis. This analysis builds on an unpublished workshop paper~\cite{schuster2023the}.

The interview recordings were transcribed and analyzed using the qualitative data analysis software Atlas.ti. Codes were created by combining deductive and inductive thematic analysis~\cite{robson2017real}. The analysis was performed by the first author of this paper, who created an initial codebook consisting of a nested coding tree with code groups, primary codes, and child codes with supporting examples. To enhance the reliability of the coding scheme, three senior researchers checked and discussed the codebook for a sample of the data until a consensus was reached. We conducted the interviews with laypeople after the expert interviews, whereby the majority of codes remained, and apart from themes specific to the interviews with the lay group (for example, concerning their news consumption habits or the felt importance of climate change for themselves) only further child codes were added.

We used an iterative data analysis process, moving back and forth between concrete conceptual codes, abstract themes, and interpretations~\cite{kekeya_analysing_2020}. The resulting codebook consisted of $4$ code groups, $44$ primary codes, and $361$ child codes. One additional code group (plus respective codes) was used for demographic data. Two separate codebooks were created for the analysis of takeaway messages. Recurring themes and the stabilizing codebook, meaning enough codes were developed to represent all important findings, were used as indicators that sufficiency of our interview data was reached in the sense that we could answer our research questions within the scope of our study setup~\cite{ladonna_beyond_2021}. For example, $90$\% of all primary codes were established after interview $18$ ($62$\% of all interviews). We stopped data collection after a total of $29$ conducted interviews. The interviews were analyzed a second time to ensure the correct assignment of codes.

\section{Findings} 
The findings are structured according to the following themes: takeaway messages by experts and lay participants (RQ-1), perceived challenges for readers (RQ-2), and design considerations for creating understandable climate change data visualizations (RQ-3). Quotes and selected findings are listed indicating group affiliation. Lay participants' statements were marked with an ID (LAY-X, as in \autoref{tab:lay-demo}) to differentiate between viewpoints; expert participants' statements are presented without ID to ensure anonymity. We do not claim the generalizability of aspects mentioned in this section; rather they represent the views in our sample. A full list of findings can be found in the supplementary material.

\subsection{RQ-1: Takeaway Messages}
\label{section:resultsmessages} 
As part of the takeaway message analysis, we analyzed the content or subjects that participants included in their messages. Participants formulated diverse takeaway messages with a high variation of included pieces of content they regarded as crucial. \autoref{tab:messagecontents} shows exemplary takeaway messages and the respective coded contents.

\begin{figure}[t]
 \centering 
 \includegraphics[width=0.65\columnwidth]{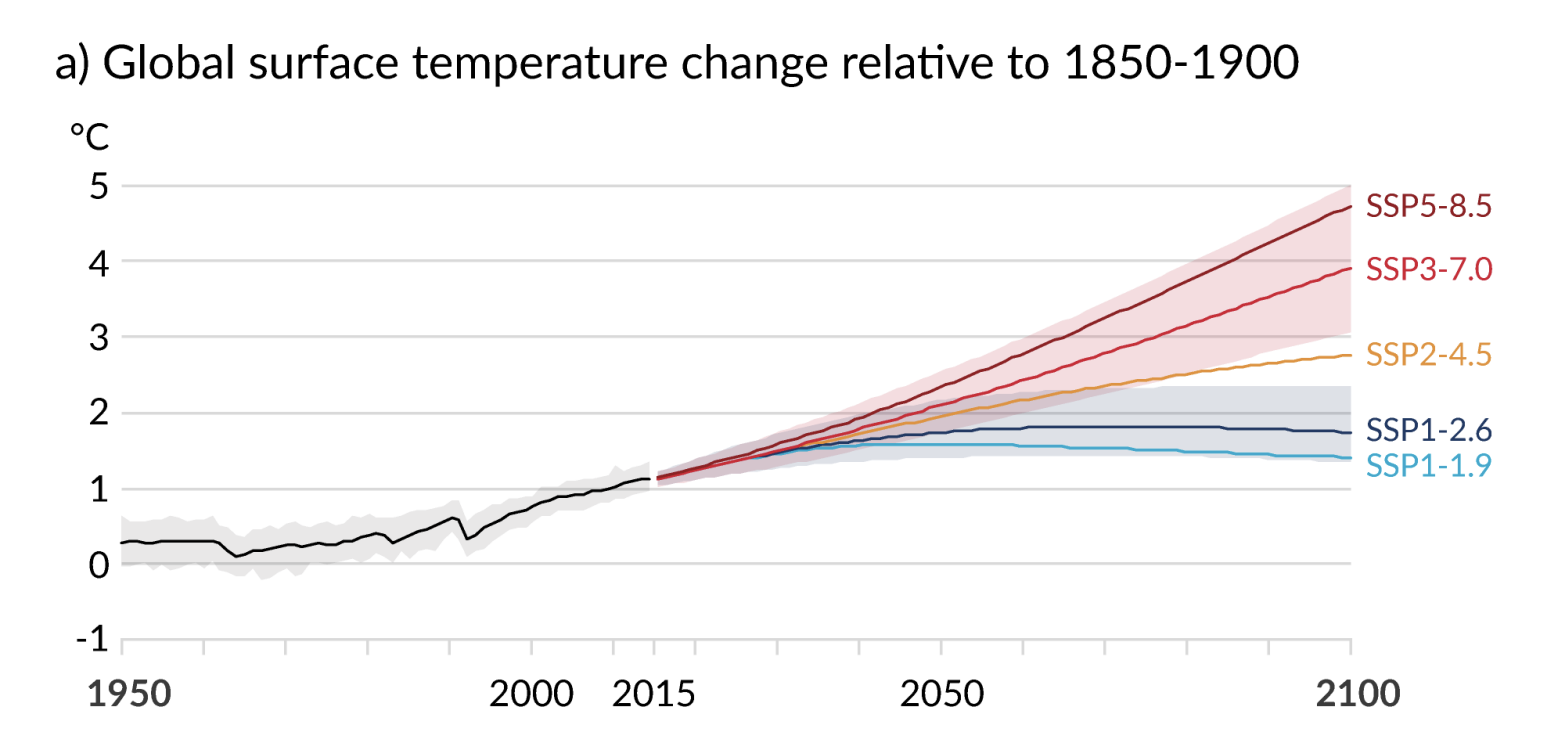}
 \caption[Caption for LOF]{Extract of Figure SPM.8 in IPCC, 2021 \cite{ipcc_summary_2021,IPCC-slides}}
 \label{fig:ipcc1}
 \vspace{4mm}
 \includegraphics[width=0.65\columnwidth]{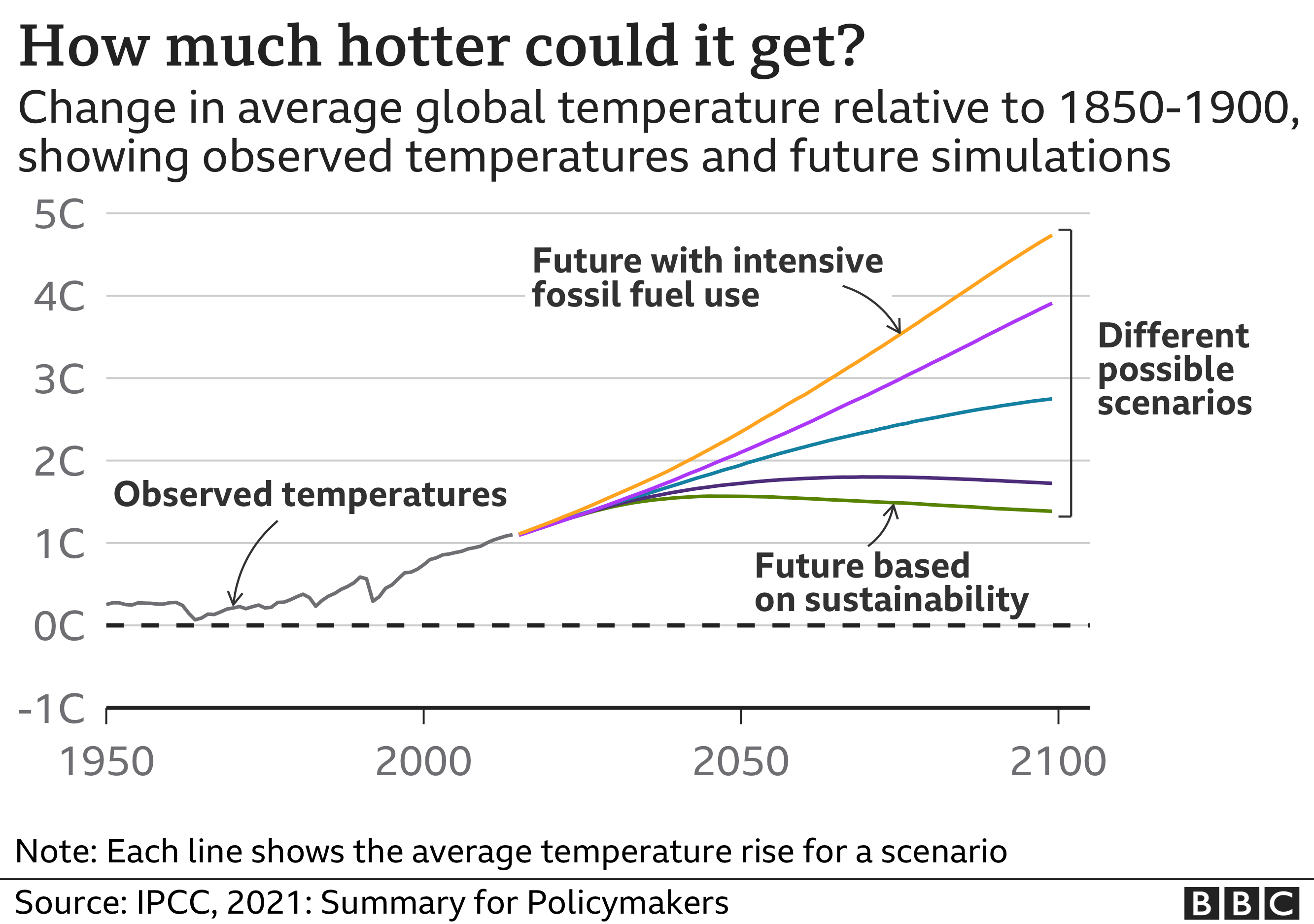}
 \caption{Visualization created by and reproduced with permission of BBC News at bbc.co.uk/news \cite{BBC}; referred to as \hyperref[fig:bbc]{Vis BBC}}
 \label{fig:bbc}
 \vspace{-3mm}
\end{figure}

\textbf{Vis BBC.} 
For \hyperref[fig:bbc]{Vis BBC}, the following subjects were included in more than half of the participants' messages: mentions of an increase, temperature, future, and positive/negative influence, or impact. Mentions of ``increase'' and ``temperature'' were coded separately, as not all participants included both of these subjects in their messages.
Some aspects were mentioned more by \textit{lay} participants than by experts, including concrete temperature specifications, specific years, increase, worst-/best-case situations, sustainability, and fossil fuel use.
Aspects like the positive or negative influence/impact of actions, the potential for improvement, and humans or mankind were more frequently included in the experts' messages. The ``relativity'' of the shown best-case scenario, as it is still getting hotter, was mentioned by $2$ \textit{lay} participants (LAY-5, LAY-10) and $1$ expert.
Messages formulated by \textit{lay} participants were considerably longer than the expert messages, with a median of $73$ vs. $29$ words per message. 
While \textit{lay} participants typically included specific details, experts tended to formulate more abstract messages: \textit{``If we focus on sustainability in the future, the global average temperature can be kept below the two-degree target that is being striven for, or even below $1.5$ degrees and then even drop slightly by $2100$. And in contrast, if we keep using everything fossil the way we have been using it, by $2100$ we will be close to five degrees of global warming.''} (LAY-7) vs. \textit{``In all cases, it's going to be hotter than it is now. And how much worse it gets depends on which scenario happens with people's actions.''} (\textit{vis expert}).

\textbf{Vis Guardian.} 
For \hyperref[fig:guardian]{Vis Guardian}, the only common contents included in more than half of the participants' messages were temperature and increase. The aspect of humans' fault was included by $7$ (of $16$) participants. 
Similarly to \hyperref[fig:bbc]{Vis BBC}, the median length of messages for this figure was more than double for \textit{lay} persons than for experts ($60$ vs. $28$ words per message).
While \textit{lay} participants mentioned the color red, the line in the chart, and concrete years more often, experts, again, tended to formulate more abstract messages: \textit{``It's clearly about the global surface temperature, again viewed over different years. And the main message is now, if I look at $2020$, for example, I see that the red line shoots up here, and steeply at that. [...] A retrospect, where it was actually, I would say, relatively the same and then a little lower, and then it got higher again, and now it is shooting up''} (LAY-5) vs. \textit{``Climate change is human-driven''} (\textit{scicomm expert}).

\textbf{Sensemaking and Understandability.} 
We asked \textit{lay} participants to voice their thinking process while seeing the visualizations for the first time, with some participants naming the chart parts they were looking at one by one. This process was highly individual, with some reading the title or subtitle before looking at the graph and others the opposite way. \textit{Lay} participants often claimed that their attention was guided by the color of the objects and mentioned bold design choices such as the prominent title of \hyperref[fig:guardian]{Vis Guardian}. Attitudes toward their ability to read data visualizations seemed to influence \textit{lay} participants' efforts, with participants who felt they are not equipped to read charts putting in less effort to understand the data. In comparison to experts, it took more time for the \textit{lay} participants to feel that they arrived at an understanding. 
While $10$ of the $12$ \textit{lay} participants appeared to interpret \hyperref[fig:bbc]{Vis BBC} in a way that matched our interpretation as researchers, LAY-1 expressed that they were having difficulties. For \hyperref[fig:guardian]{Vis Guardian} $5$ \textit{lay} participants appeared to interpret the graph in a way that is supported by the shown data; the interpretation of $5$ further \textit{lay} participants was unclear and LAY-1, LAY-3, LAY-6, and LAY-8 expressed having difficulties (the $2$ latter did not formulate a message). However, not only \textit{lay} participants had difficulties interpreting the chart, but also $2$ \textit{scicomm} experts stated that they struggled to deduct what the chart should tell them. \autoref{tab:lay-demo} shows the \textit{lay} participants' opinions towards the visualizations.

\begin{table*}[tb]
\mysize
\caption{Example messages and coded contents}
\centering
\label{tab:messagecontents}%
\begin{tabular}{p{1.34\columnwidth}p{0.63\columnwidth}}

    \hline
    \vspace{0.01mm}
    \textbf{Example messages} & \vspace{0.01mm} \textbf{Coded contents} \\
    \midrule 
    
    ``The effect that sustainability could have on the temperature or global temperature on average. And there are like 1, 2, 3, 4, 5 scenarios. So I think the main message is that sustainability would help. And that, if you continue to live like we do it, we would be at plus five, I guess, degree Celsius.'' (LAY-3 about \hyperref[fig:bbc]{Vis BBC}) 
    & Sustainability, positive/negative influence/impact, global earth average, temperature, scenarios, 5-degrees \\
    \multicolumn{1}{c}{ }&\multicolumn{1}{c}{ }\\[-0.2cm]
    
    ``They show how different the future can be, depending on humans, how humans behave.'' (\textit{vis expert} about \hyperref[fig:bbc]{Vis BBC}) 
    & Positive/negative influence/impact, future, human influence/fault \\
    \multicolumn{1}{c}{ }&\multicolumn{1}{c}{ }\\[-0.2cm]

    ``Things are going extremely up now. Like the red line, from 1850, it was still down, and then in 2020, we are already a long way up.'' (LAY-12 about \hyperref[fig:guardian]{Vis Guardian})  
    & Increase, severity, red, line, year-1850, year-2020, difference \\
    \multicolumn{1}{c}{ }&\multicolumn{1}{c}{ }\\[-0.2cm]

    ``We see that in a very short time that the earth got hotter. So there's no way this is a normal cycle if you compare it to the historic data. So it's us basically, is what the message is.'' (\textit{vis expert} about \hyperref[fig:guardian]{Vis Guardian})
    & Short time frame, global earth average, temperature, increase, anomaly, past, human influence/fault \\
    
    \hline
\end{tabular}
\vspace{-3mm}
\end{table*}

\footnotetext{Mentioned channels: Instagram, Facebook, Linked-in, Twitter}

\subsection{RQ-2: Common Challenges}
\label{section:resultsdifficulties}
\begin{figure*}[t]
\centering
\begin{minipage}{.28\textwidth}
  \centering
  \includegraphics[width=\linewidth]{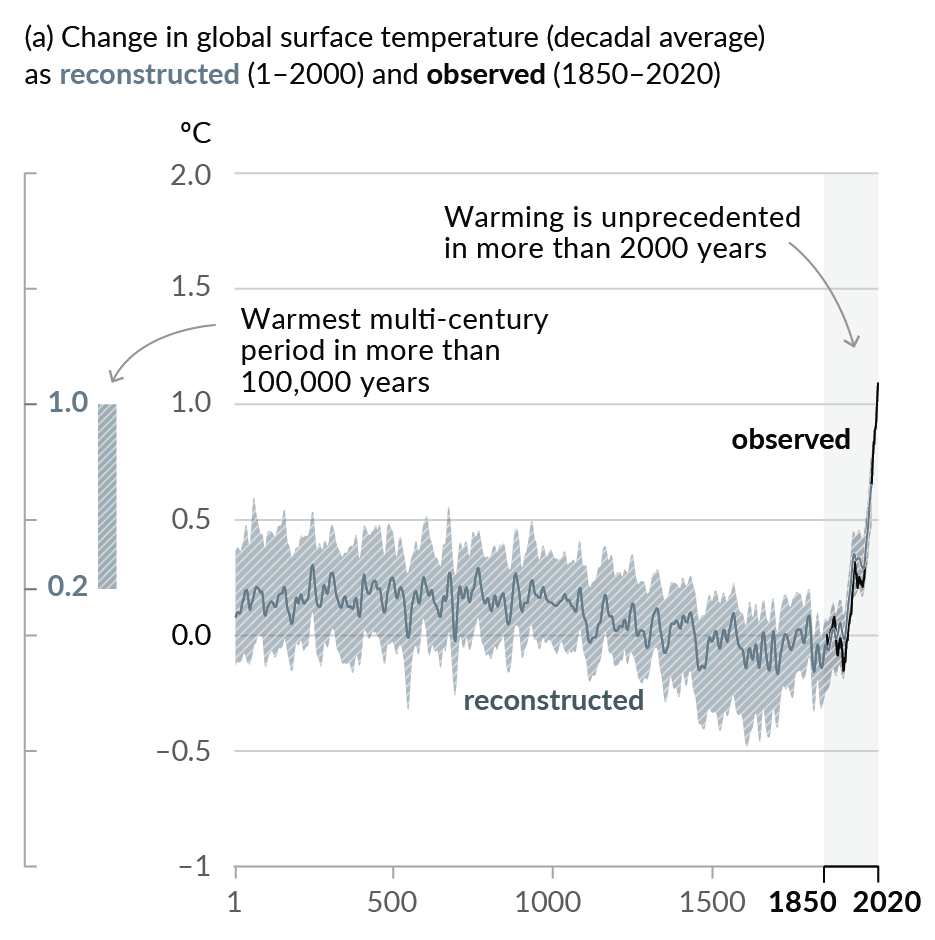}
  \captionof{figure}{Extract of Figure SPM.1 in IPCC, 2021 \cite{ipcc_summary_2021,IPCC-slides}}
  \label{fig:ipcc2}
\end{minipage}
\hspace{0.7cm}
\begin{minipage}{.28\textwidth}
  \centering
  \includegraphics[width=0.9\linewidth]{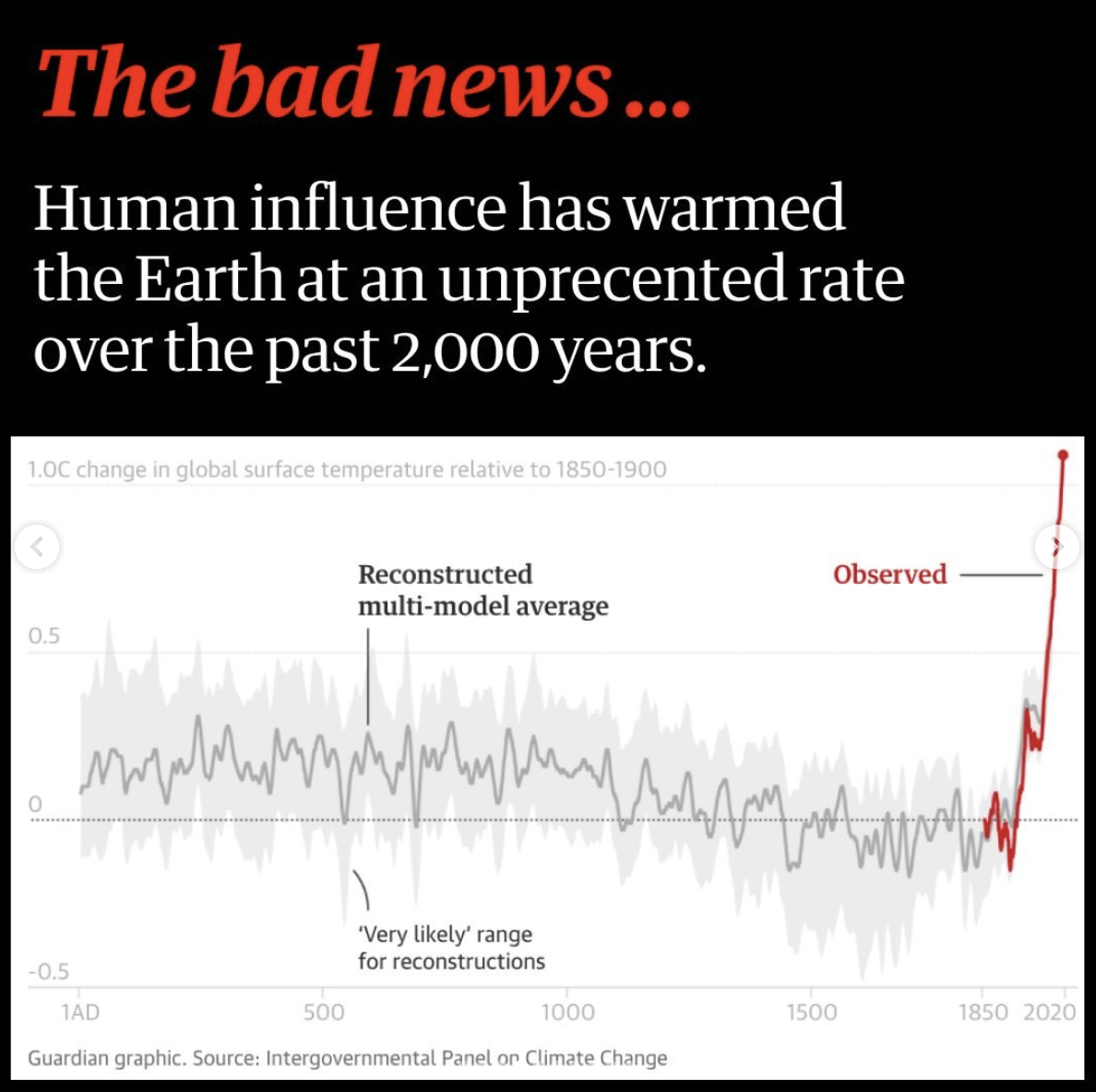}
  \captionof{figure}{Visualization created by and reproduced with permission of Guardian News \& Media Ltd 2023 \cite{the_guardian_as_2021}; referred to as \hyperref[fig:guardian]{Vis Guardian}}
  \label{fig:guardian}
\end{minipage}%
\hspace{0.7cm}
\begin{minipage}{.28\textwidth}
  \centering
  \includegraphics[width=\linewidth]{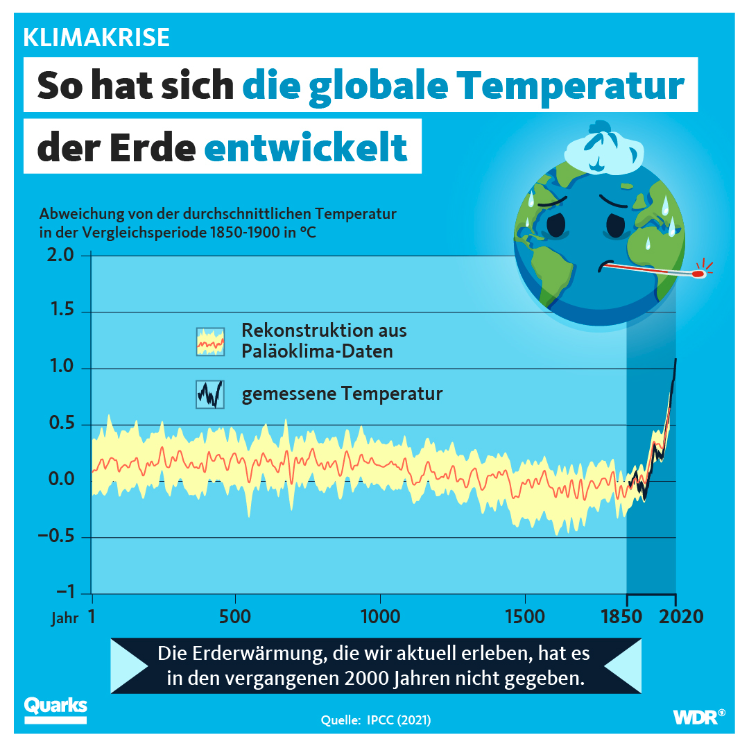}
  \captionof{figure}{Visualization created by and reproduced with permission of WDR \cite{noauthor_quarks_2022}}
  \label{fig:quarks}
\end{minipage}%
\vspace{-3mm}
\end{figure*}

Participants mentioned various benefits of visual communication including the potential to increase attention and engagement, complexity reduction, and visual preference. $11$ experts and $8$ \textit{lay} participants stated that they would first look at a data visualization in a news article before reading the text. However, experts from all groups acknowledged that interpreting data visualizations about climate change can be difficult for laypeople. The opinions about the understandability of the example figures were quite diverse: 6 experts and $5$ \textit{lay} interviewees thought \hyperref[fig:bbc]{Vis BBC} would be difficult to interpret for many people; $5$ experts and $6$ \textit{lay} participants thought so for \hyperref[fig:guardian]{Vis Guardian}.

\textbf{Vis BBC.} 
While \hyperref[fig:bbc]{Vis BBC} seemed to be easier for participants to grasp than the second visualization, $6$ interviewees commented that the missing degree Celsius symbol on the temperature axis slowed down their sensemaking process. With $8$ participants speaking in favor of the title and the annotations, $12$ interviewees from all groups criticized that the used wording is too complicated for a news source: \textit{``It's too complex wording: `different possible scenarios', [...] `future based on sustainability'. Nice for us scientists, for the broader public: way too complicated''} (\textit{climate expert}). In regard to the subtitle, $4$ \textit{lay} participants acknowledged that they would not understand why the 1850-1900 time frame was chosen as a reference. LAY-1 seemed to misinterpret some connections: \textit{``This fuel and sustainability actually have nothing to do with the rise in temperature and anything like that.''} (LAY-1).

\textbf{Vis Guardian.} 
Participants felt less confident about their interpretation of \hyperref[fig:guardian]{Vis Guardian}, partially mentioning the graph fluctuation, which reminded them of a cardiogram or a lie detector test, as a reason for the felt increase in complexity. Wording like ``reconstructed multi-model average'' and ```very likely' range for reconstructions'' was an issue for $7$ \textit{lay} participants. Again, 4 participants expressed that they would not be able to understand the reference time frame without further explanation. The chosen time axis was the reason for much discussion: LAY-8 and LAY-9 expressed doubts about the unprecedented rate of the temperature increase, claiming that it would be necessary to look at a larger time frame than just the last 2020 years, also after looking at the original version by the IPCC (\autoref{fig:ipcc2}). LAY-10 criticized the time frame as too large, as humans did not record temperature data for such a long time and hence doubted the correctness of the data. Such clarity issues resulted in $3$ \textit{lay} participants strongly doubting the information, although their voiced attitudes towards the topic would not have indicated a general skepticism.

\textbf{Interpreting Data Visualizations Is Difficult.} 
Participants thought that interpreting data visualizations requires different kinds of background knowledge. Areas of education that should be emphasized include mathematics, visualization knowledge, and basics about predictions and uncertainty. Experts mentioned the lack of visual data literacy, or graphicacy, among the public as a problem: \textit{``One mistake we in the visualization community make is that we overestimate the visual literacy of people''} (\textit{vis expert}). Participants shared their opinions about common difficulties and named uncommon visualization types as one of the main factors that hinder understanding. \textit{Vis} and \textit{scicomm} experts mentioned that colors can be the cause of misinterpretation due to different cultural or emotional connotations. Predictions, uncertainties, as well as scales/axes were said to be difficult for lay users: \textit{``You have to teach this [...], what is a prediction about, what is the confidence interval [...], nobody knows this''} (\textit{scicomm expert}).

\textbf{Climate Change Is Complex.} 
On a scale from $1$ (low) to $10$ (high), all but $1$ \textit{climate} expert rated the understanding among the public concerning climate change and its consequences below or equal to $5$. \textit{Lay} participants named insufficient understanding due to the complexity of climate change and insufficient education as causes for the general lack of action. Both of these aspects were also mentioned by experts as major factors why interpreting data visualizations can be difficult: \textit{``It's just a very complex topic and unless you're willing to spend serious time on thinking about it, I believe it's just hard''} (\textit{climate expert}). Experts also acknowledged that a deep understanding among the public might not be a realistic goal and it is more the trust in science that gets laypeople to take action: \textit{``You have to trust science, which is very close to believing in something because you cannot reproduce that knowledge''} (\textit{scicomm expert}).

\subsection{RQ-3: Design Considerations}
\label{section:resultsimplicationsaudience}
The importance of the perceived quality of a visualization was mentioned by participants among all expert groups. However, creating high-quality data visualizations is not a trivial task: \textit{``It's very hard to create a graphic that is unambiguous in its message. Can be done, but requires work. [...] It's a technical issue and thus can be taught. And we don't do that''} (\textit{climate expert}). Participants shared their opinions on what makes a high-quality visual for them and made recommendations on how to create effective climate change data visualizations (RQ-3). We summarized the findings under three main themes: Embrace the audience; make it simple, but transparent; and make it attractive.

\subsubsection{Embrace the Audience} 
According to the experts, the assessment of whether a visualization is well-designed or not strongly depends on the visualization's purpose, outlet, or audience: \textit{``I argue to students, there's not a good or a bad visualization — it really depends on the purpose''} (\textit{vis expert}). Adapting climate change data visuals to the knowledge and needs of the users was named by \textit{vis} and \textit{scicomm} experts as a way to influence the quality of data visualizations. 

\textbf{Relatability of Contents.} Participants have emphasized that relatability to the contents of a data visualization plays a big role in getting and keeping the readers' attention: \textit{``You really have to make sure that people can relate with whatever you communicate [...], which is definitely difficult when it comes to climate change because it's a very abstract thing''} (\textit{scicomm expert}). While \textit{vis}/\textit{scicomm} experts have mentioned creating relatable visuals as a general recommendation, $9$ participants have criticized \hyperref[fig:bbc]{Vis BBC} for not showing any consequences. Visualization types were mentioned as a means to increase relatability, for example, maps to show the consequences for specific regions. \textit{Lay} participants claimed that climate change content that focuses on the future, showing consequences or solution strategies is most relatable. 

\textbf{Designing for Different Channels.}
According to participants, requirements for creating data visualizations massively differ between channels such as online vs. print, static vs. video, or article vs. social media post:  \textit{``I can complain that a visualization is far too simple and doesn't show enough detail. Well, if you have three seconds to look at the picture, it's good if it's very simple''} (\textit{vis expert}). LAY-2 and LAY-3, who use Instagram as their primary source of information, explained that content needs to adhere to different standards on social media: Instagram would be used for pleasure and users might not stop scrolling for a complicated graphic. They suggested a format by the German Instagram channel ``Tagesschau'' \cite{noauthor_tagesschau_nodate}: using the slideshow function, they typically include several pieces of content, starting with something lightweight like a photograph and a catch-line. Consecutively, users can then swipe through the slides to learn more in small digestible doses, including bullet points, photographs, videos, quotes, or charts. LAY-10 claimed that they would prefer short explanatory videos including animated visualizations that build up over time.

\textbf{Designing for Reachable Audiences.} Participants reflected on different target groups within the general public. LAY-9 explained that in their opinion the public can be split into three groups: people who are already convinced that climate change is a problem, people who will never believe in climate change, and people, who have no strong opinion. This third group might be reached through infographics about the potential impact of personal decisions. However, visualizations could easily have the opposite effect, as contrarians, people who doubt predominant science about climate change, might be skeptical about \hyperref[fig:bbc]{Vis BBC} as it is not concrete enough. The importance of clarity of the displayed contents was mentioned by several participants. Leaving any ambiguity open for interpretation (like whether or not there were times with similar temperatures before in \hyperref[fig:guardian]{Vis Guardian}), could result in critical readers not trusting the visualization. 

\textbf{Human-centered Process.} In the context of creating visuals for their intended target audience, experts mentioned a user-centered design process: factors included co-design, iterative design, and framework conditions such as time, funding, and expertise. An iterative co-design process was described as the collaboration between climate change scientists and visualization experts (with communication mentioned as a challenge) that starts with the purpose of a visual and passes through several design iterations: \textit{``Embrace the audience in the design process to make sure that you understand how they interpret the information''} (\textit{scicomm expert}).

\subsubsection{Make it Simple, but Transparent} 
\label{section:resultsimplicationssimplicity}
Participants among all groups named simplicity and complexity reduction as key factors in creating effective data visualizations: \textit{``If the visualization is done very good, it should be super easy to understand it. And that's the problem with most visualizations, [they] are still too complex''} (\textit{climate expert}). 

\textbf{Clear Intent.} Experts mentioned that the definition of a clear message or intent can positively influence understandability: \textit{``If you don't know what the intent is, [...] there's no way you can ever do a good graphic, except by complete accident''} (\textit{climate expert}). While the ambiguity about the scenarios in \hyperref[fig:bbc]{Vis BBC} was brought up by $12$ participants, some explicitly criticized the main point the figure seems to make. They were missing more quantitative details about the scenarios and what kind of behavior change would be required for each of them: \textit{``What does that mean for my life? [...] It's always good to have some recommendations for actions''} (LAY-4). One \textit{scicomm} expert pointed out that the visualization's purpose appears not to be about giving precise numbers, but only wants to show that there is \textit{``some room for taking levers to have an impact on the future''} (\textit{scicomm expert}).  

\textbf{Simplification.} The issue of simplification vs. accuracy was discussed throughout the interviews, with a \textit{climate} expert naming ``being simple on complex issues'' as one of the most difficult challenges. Factors that support this need for simplification include the complexity of climate change and readers' missing time/involvement: \textit{``The risk you run, if you insist on scientific rigor [...], is that you might lose your audience altogether. And then you're not conveying the depth of your message either''} (\textit{climate expert}). 
While the simplifications made in \hyperref[fig:bbc]{Vis BBC} compared to \autoref{fig:ipcc1} were mentioned as positive, some participants thought that it was still too complicated due to the overall complexity ($8$ participants), the wording ($12$ participants) or the visualization type ($6$ participants): \textit{``This would not be the type of data visualization I would envision for the broader public. It is too complex''} (\textit{climate expert}). Contrarily, LAY-9 criticized that \hyperref[fig:bbc]{Vis BBC} did not go far enough and that in the discussion of climate change a stronger scientific presence is needed. Some experts also mentioned over-simplification (such as excluding uncertainty ranges) as a problem, with \textit{scicomm} experts being more set on maintaining scientific accuracy. However, most participants favored depicting climate change content as simply as possible and putting additional information in the text or citing more detailed sources.  

\textbf{Uncertainty.} With \textit{climate}/\textit{vis} experts naming visualizing uncertainty as a major challenge, opinions on whether to include or exclude uncertainty in climate change visuals for the public varied among participants. For \hyperref[fig:bbc]{Vis BBC}, $8$ experts and $8$ \textit{lay} participants regarded it as positive, and $3$ \textit{climate}/\textit{scicomm} as negative that the uncertainty ranges of \autoref{fig:ipcc1} were not shown in the BBC's version. Independent of the examples, $5$ \textit{climate}/\textit{scicomm} experts spoke in favor of showing uncertainty to not lose the reader's trust; \textit{vis} experts argued for excluding the uncertainty for lay audiences unless it eliminates the heart of the matter: \textit{``I don't see an added benefit of putting that burden of uncertainty onto the [general public] because it doesn't make a huge difference for the main message''} (\textit{vis expert}). 
For the uncertainty ranges shown in \autoref{fig:ipcc1}, $5$ \textit{lay} participants either had trouble understanding them themselves or thought that it would be hard for others; $7$ \textit{lay} participants did not mind or preferred the uncertainty in \autoref{fig:ipcc1}: \textit{``I'd rather know the uncertainty range. And then again, how important is it?''} (LAY-4). LAY-7 spoke in favor of the uncertainty ranges to ease the public's concern (despite the deviations in both directions).

\subsubsection{Make it Attractive} 
\label{section:resultsimplicationsattractiveness}
Discussing the understandability of \hyperref[fig:bbc]{Vis BBC}, LAY-3 emphasized: \textit{``The main problem is: would you actually look at it?''} (LAY-3). \textit{Climate}/\textit{scicomm} experts have also mentioned the lack of attention as one of the main problems in visual data communication. We asked \textit{lay} participants if the example visualizations would catch their attention if they saw it in their everyday news sources. For \hyperref[fig:bbc]{Vis BBC}, $6$ \textit{lay} interviewees stated that it would catch their attention; $7$ found \hyperref[fig:guardian]{Vis Guardian} engaging: \textit{``It catches your eye and you can't in good conscience just go on spending your time on Instagram without knowing what the bad news is''} (LAY-2). However, the attention-getting design was not to everybody's liking: \textit{``That's like an accident, you look there to see if there's blood''} (LAY-8). 

\textbf{Design Choices.} Principles of good design and aesthetics were mentioned by participants from all groups: The amount of information should not overload the graphic, contents should be presented clearly, and user perception should be guided through positioning and highlighting. \hyperref[fig:bbc]{Vis BBC} appeared to be cluttered for $10$ interviewees, who suggested a reduction of text or different positioning of the annotations. Participants had different opinions concerning the line coloring in \hyperref[fig:bbc]{Vis BBC}. While $3$ \textit{vis} experts spoke in favor of the categorical colors, $7$ experts and $8$ \textit{lay} interviewees felt that the colors seem random and would have liked a blue-to-red color scale like in \autoref{fig:ipcc1}: \textit{``The color for the worst case is completely wrong, it should actually be deep red and not somehow yellow; maybe the air will be yellow then''} (LAY-10). 
Design choices also highly affected perceived trustworthiness. All \textit{lay} participants, except LAY-7 and LAY-8, found \hyperref[fig:bbc]{Vis BBC} trustworthy. 
In contrast to that, opinions on the trustworthiness of \hyperref[fig:guardian]{Vis Guardian} were split, with LAY-2 perceiving the figure as a neutral communication of facts and LAY-4 calling the design a \textit{``crime against humanity''} because of its emotional framing and dark and judgmental design. This was the reason why $4$ further participants felt manipulated (\autoref{tab:lay-demo}). LAY-7 perceived the scales to be incorrect (even though similar to \autoref{fig:ipcc2}): \textit{``The scientific graphic is true to scale [...], which is not the case for [\hyperref[fig:bbc]{Vis BBC}]. It was stretched insanely: the good part from the old days and the bad part, for which we are responsible''} (LAY-7).

\textbf{Infographics.} Infographics were mentioned by \textit{vis}, \textit{scicomm}, and \textit{lay} participants as a way to engage the general public in complex topics. \hyperref[fig:bbc]{Vis BBC} was criticized by $2$ \textit{scicomm} experts and $1$ \textit{lay} participant for its lack of pictures/icons. We realized during the interviews that \textit{lay} participants partially lacked an understanding of the term infographic. We hence brought an example that shows the same data as \hyperref[fig:guardian]{Vis Guardian} in a more casual style (\autoref{fig:quarks} to the last $6$ interviews. None of the \textit{lay} participants felt that this design was less trustworthy. LAY-6 has not enjoyed any of the previously shown visualizations and felt unable to formulate a takeaway message for \hyperref[fig:guardian]{Vis Guardian}: \textit{``This looks so technical, so highbrow, so professional. [...] I would think to myself; I don't need to look; I don't get it anyway''} (LAY-6). However, when shown the infographic version, they expressed their enthusiasm and shared that they feel that even children should now get the main message. LAY-10, while not opposed to the idea of infographics, expressed that the content is still too complicated for anyone to understand, while the design looks like it was made for toddlers.

\textbf{Interaction.} Participants stated that interactive visualization techniques can be a powerful way to engage users. While interactivity can be suitable for specific purposes, such as for historical data, participants argued that creators should not overuse it: \textit{``Interactivity has to be a consequence of [the user's] needs. It's not just something that we decide to do because it's cool''} (\textit{scicomm expert}). Mentioned benefits and drawbacks include: increased engagement (possibility to explore data), increased relatedness (possibility to pose own questions), limited accessibility (technical means necessary), and limited engagement (interest can drop quickly). While \textit{lay} participants confirmed the potential relatability benefits, none of them reported having used an interactive tool.

\begin{table*}[tb]
\footnotesize
\caption{Overview of key themes and findings during the interviews and comparison to previous work}
\centering
\label{tab:findings}%
\rowcolors{2}{white}{gray!15}
\renewcommand{\arraystretch}{1.5}
\begin{tabular}{p{0.22\columnwidth}p{1.2\columnwidth}p{0.44\columnwidth}}

    \hline
    \textbf{Theme} & \textbf{Findings during interviews (partially comparing \textit{lay} and expert participants)}  & \textbf{Comparison to previous work}\\  \midrule 

   \textbf{Attention} & 
   $\bullet$ Perceived attractiveness of a data visualization depends on channel \newline
   $\bullet$ Willingness to look at a data visualization depends on perceived simplicity \newline 
   $\bullet$ \textit{Lay} participants evaluated bold design choices as more attention-grabbing but less trustworthy &
    Validating \& extending prior work on IPCC graphics ~\cite{harold_enhancing_2017,harold_communication_2020} and general design principles~\cite{williams_non-designers_2017} \\

    \textbf{Sensemaking} & 
    $\bullet$ The process was influenced by a person's prior knowledge, attitudes, and experiences \newline 
    $\bullet$ \textit{Lay} participants' attention was guided by lower-level design details, such as color, lines, text  & 
    Partially supported by previous studies like~\cite{peck_personal_2019,hullman_visualization_2011,borkin_memorability_2016,kong_frames_2018} \\

    \textbf{Understandability} & 
    $\bullet$ Most participants spoke in favor of simplifications \& using text for further explanations \newline
    $\bullet$ Participants criticized the visualizations for their complex terminology/lack of explanations \newline
    $\bullet$ The majority of participants preferred leaving out the uncertainty for the general public \newline
    $\bullet$ Experts claimed that the visual literacy of the public is commonly overestimated \newline
    $\bullet$ \textit{Lay} participants found infographics more understandable and not less trustworthy & 
    Extending and supporting previous work by~\cite{kause_visualizations_2020, harold_cognitive_2016, corner_principles_2018}; challenging findings about infographics' negative impact on credibility~\cite{mcmahon_scientific_2016} \\

    \textbf{Takeaway \newline Messages} & 
    $\bullet$ The median length of formulated messages was more than double for \textit{lay} participants \newline 
    $\bullet$ Messages formulated by \textit{lay} participants were less abstract and contained lower-level details &
    Previous work focuses on different aspects \\
    
    \textbf{Action} & 
    $\bullet$ Participants asked for relatable content (consequences of climate change or individual actions) \newline
    $\bullet$ Especially \textit{lay} participants were opposed to content that is negatively framed & 
    Partially confirming previous research~\cite{corner_principles_2018,ferreira2021Climate} \\
    
    \hline
\end{tabular}
\vspace{-3mm}
\end{table*}

\section{Discussion} 
 With previous studies in the field mainly investigating IPCC visualizations and their target audience, our study provides insights into which factors are important in formats for the general public. We build on and validate past work, but we also extend it as we i) focus on news sources, ii) combine perspectives of experts in different fields and laypeople, and iii) shed light on what the participants read from a visualization through investigating takeaway messages. Our study shows how experts and laypeople emphasize specific aspects and have different beliefs and opinions. While partially being a factor of people's individual expertise and experience, it also highlights the contextual nature of understanding data and visualizations \cite{burns2020how}. This section summarizes the interviews' most prevalent findings and situates them in existing literature (\autoref{tab:findings}).

\textbf{Designing for Attention.} 
Prior work by Harold et al.~\cite{harold_enhancing_2017} gives recommendations to IPCC researchers for creating effective visuals, which partially align with our findings. However, previous studies make their claims about the IPCC target audience and outreach through IPCC channels. In our interviews, it became apparent that scientific sources like the IPCC are not something that laypeople typically see. Rather, we must think about climate change communication happening at the front lines of news or social media channels. With the abundance of information provided to us, the question of whether or not users would look at a visualization constitutes the first hurdle. An important factor in attracting users' attention according to the interviews and previous work is the reduction of perceived complexity in a visual~\cite{williams_non-designers_2017}. While participants provided suggestions on how to make climate change data more ``digestible'' (such as using Instagram slides and mixing up contents), it also became clear that design choices have to be carefully contemplated. Dramatic designs emphasizing human fault seem to increase attention but potentially lower trustworthiness, as readers feel manipulated. More studies are necessary to determine how we can effectively design for different groups of the general public and what kind of data visualizations are capable of acquiring and keeping users' attention in real-life scenarios.

\textbf{The Individuality of Sensemaking.} 
Our study confirms findings from prior research (such as~\cite{rothermich_influence_2021, peck_personal_2019, Deng2016}) suggesting that the process of understanding data visualizations is highly individual and influenced by a person's prior knowledge or education, attitudes, beliefs, and experiences. \textit{Lay} participants who felt they lacked the skills to understand data visualizations made less of an attempt to interact with the data; their beliefs about the trustworthiness of news media sources influenced whether they liked and trusted the example visualizations. Both, perceived understandability and trustworthiness, were also heavily influenced by the visualizations' design choices.
We saw similarities to the findings of Lee et al. \cite{lee_how_2016}, who have categorized novices' visualization sensemaking along five cognitive activities. While we have encountered instances of those activities during the interviews with \textit{lay} participants, the occurrence or sequence of those were, again, highly individual.
Contrary to the experts, \textit{lay} participants' attention was highly influenced by the color of objects and textual elements, with the formulation of the title being crucial for their interpretation. 
Those findings are partially coherent with previous research stressing the importance of the narrative of visuals~\cite{hullman_visualization_2011, fish_storytelling_2020}, as well as users' focus on the title~\cite{borkin_memorability_2016}. Kong et al.~\cite{kong_frames_2018} have shown that the framing of a visualization title can influence the perceived main message and warned about viewers’ unwarranted conviction of the neutrality of visualizations. For our specific sample of laypersons, it was striking how many doubted the shown information, especially for \hyperref[fig:guardian]{Vis Guardian}, which they perceived to have a very sensational title.
We conclude that design choices play a crucial role when communicating with laypeople, much more so if the topic is as emotionally loaded as climate change.

\textbf{Designing for Understandability.} 
Our findings suggest that substantial simplifications in visualizations for the public, for example, compared to IPCC visualizations, are necessary to meet the needs of public audiences. However, visualization designers must be transparent to avoid oversimplifying content and potentially losing the reader's trust. In this regard, text can play an important role. When a data visualization excludes information for the sake of simplicity, interviewees suggested that accompanying text could bridge the gap between simplification and accuracy. A major point of criticism regarding the example visualizations has been the perceived complexity (in general and concerning wording). The reduction of complexity for climate change visuals for the public was recommended previously~\cite{harold_cognitive_2016}, and the usage of simple captions and known terminology was shown to have a positive impact on readers' understanding~\cite{kause_visualizations_2020}. 
Our findings provide insights into the understandability of common climate change line charts in news sources, suggesting that the level of complexity shown in the example visualizations was still too high for some participants and would require explanations in the text. This is supported by expert opinions about public visual literacy, stressing that the visualization community tends to overestimate the visual literacy of laypersons and climate change content should be put even simpler to reach diverse audiences. This issue seems to result from difficulties reading visualizations, like not feeling equipped to read a chart, but also from complexities of the topic itself, such as not understanding specific wording or details regarding the underlying models (such as the reference time frame in the example visualizations). Summarized, we find that visualization literacy cannot be discussed solely as a function of the chart but needs to be interpreted in the context of the topic, the data, and, if relevant, the underlying model, all in light of the envisioned audience and their individual differences and backgrounds.

While the difficulty of understanding uncertainty in public climate change communication was also acknowledged previously~\cite{corner_principles_2018}, we specifically asked if showing uncertainty ranges in public communication is necessary or counterproductive. Our sample predominantly spoke in favor of not showing the uncertainty ranges in a visual for a newspaper source to reduce complexity and instead providing explanations about it in the accompanying text. Mentioned solutions that are possible in an online or video context are adaptable visualizations or visual storytelling that builds up complexity step by step, allowing people to stop engaging when they see fit, as can be seen in the work of some practitioners. 

Participants acknowledged the benefits of incorporating images and icons into climate change visuals, as they appear more interesting and understandable. Previous research has shown that infographics can increase user engagement and aesthetic value~\cite{franconeri_science_2021}, which aligns with the assessments of the \textit{lay} participants. McMahon et al.~\cite{mcmahon_scientific_2016} investigated the perception of IPCC graphics in comparison to infographics, claiming that infographics can increase accessibility but might have a negative impact on scientific credibility. For our sample, the infographic (\autoref{fig:quarks}) was indeed perceived as more accessible but not as less trustworthy. Future research could shed light on how data visualizations about climate change are created and used in news media sources and test how different design options (such as including or excluding uncertainty, infographic style) affect readers' interest and understanding.

\textbf{Takeaway Messages on Different Levels.} 
By investigating takeaway messages of two exemplary visualizations, we observed differences in the included contents among and within participant groups. \textit{Lay} participants' messages were shorter than the expert ones due to different levels of abstraction. This shows similarities to findings in communication research in which experts tended to use more abstract instructions than beginners when providing task instructions~\cite{hinds_bothered_2001}. In the context of natural language descriptions of visualizations, Lundgard and Satyanarayan~\cite{lundgard_accessible_2021} introduced four levels of semantic content, namely $1$) visualization construction properties, $2$) statistical concepts/relations, $3$) perceptual/cognitive phenomena, and $4$) domain-specific insights. The contents of our takeaway messages can be categorized along those levels: \textit{lay} interviewees more frequently included lower-level contents like visualization properties or statistical details (levels $1$ and $2$); experts tended to formulate messages on a higher level, often referring to conclusions, trends, and contexts (levels $3$ and $4$). 
A possible reason might be that experts are more likely to have seen similar charts or are familiar with interpreting data visualizations in general. Hence, visualization properties might not stand out to them in a way that they would include them in their takeaway messages. 
On top of the expertise, also the visualization design seemed to influence the usage of lower vs. higher-level content: visualization properties, like colors or lines, were more often mentioned in messages for \hyperref[fig:guardian]{Vis Guardian} than for \hyperref[fig:bbc]{Vis BBC}. 
It is interesting how different aspects are essential for different readers, which might eventually shape what we remember about a visualization~\cite{bateman_useful_2010}. Future work could explore further influencing factors on sensemaking and message formulation, as well as how readers remember data visualizations based on their individual takeaways.

\textbf{From Visualization to Action.}  
How do we design climate change data visualizations that adhere to the needs of diverse users? Literature and our findings indicate that a highly iterative, user-centered co-design process, including domain and visualization experts as well as user-testing, is the way forward \cite{morelli_co-designing_2021}. While such approaches have been in place for IPCC visualizations, experts suggested that these practices can also benefit news sources. In general, our findings show parallels to principles of science communication (for example \cite{fischhoff_sciences_2013}) and visual data communication (for example \cite{franconeri_science_2021}). However, what makes climate change stand out is the complexity of the data, the urgency to act, and the emotional aspect that design choices can trigger. The feeling of responsibility for the current situation, fear about the future, and the perceived need or social pressure to change personal behavior were apparent during the discussions. Especially, acknowledging the need for personal behavior change is inconvenient for many people. Hence, not only a lack of interest in and understanding of climate change but also cognitive phenomena, like the status quo bias (like in \cite{samuelson_status_1988,patt_action_2000}), could restrain us from giving ``climate action the high priority it is due'' \cite[p. 3]{boehm_state_2021}. 
The interviews suggest that an important factor in motivating users to engage with and act according to a data visualization lies in the relatability of the contents, which is, in part, comparable to Corner et al.'s \cite{corner_principles_2018} principles for public engagement. Our participants shared concrete suggestions on addressing users' interests, including showing the consequences of climate change on a map \cite{fish_cartographic_202}, visualizing the impact of actions, and giving concrete recommendations for action. Building on previous work \cite{ferreira2021Climate}, our findings also show that messages are most effectively framed in a positive solution-oriented way, which motivates readers to act.

\textbf{Reflections about Studying Laypersons.} 
Burns et al.'s \cite{burns_novices_2023} recent work on visualization novices is a helpful resource for thinking about laypeople. Despite the justified concerns of defining laypersons in terms of their perceived limitations in comparison to an expert group, we were still in need of a definition that excluded persons from the \textit{lay} group who had a professional background in our chosen areas of expertise. As we were interested in people's opinions with different habits, experiences, and interests, we decided against limiting our \textit{lay} sample in other ways, such as news or visualization consumption. While some participants reported seeing data visualizations on a regular basis, others would barely encounter any graphics in their day-to-day life. However, it was valuable to also capture their perspectives, for example, on what would catch their attention and what they would find understandable. We hence had to significantly adapt how we ask questions and talk about `data visualizations', which in itself was not an easy-to-understand term for many. We also did not want to create an atmosphere where participants felt that their knowledge or understanding of a specific visualization was being tested or judged. We realized that this feeling was mainly present in the questions about the main takeaway message of the visualizations. We explained in detail the aims of this study to make participants feel comfortable sharing their thought processes and opinions.

\textbf{Limitations.} 
Our findings have to be understood within the limits of the study design. Despite efforts to interview participants with diverse backgrounds, the interviews are still limited in number in terms of fields of expertise and backgrounds. Our sample shows a focus on European and German-speaking participants, with some of them making their arguments for local circumstances. We identified important themes for public climate change communication, but further studies with a larger and more diverse (culturally and geographically) sample are necessary to test concrete assumptions. For the example visualizations, we opted for line charts as one of the most common choices of IPCC authors \cite{wozniak_stakeholders_2020} and for our sample of news article visualizations. While the example visualizations were very helpful in encouraging a discussion and also led to participants sharing their general recommendations, both visualization pairs depicted line graphs with a time axis and temperature trends. We acknowledge that this choice shaped our findings, which could be accounted for with different chart types in follow-up studies. Finally, our questions were aimed at opinions about broader themes and findings might, in part, not be specific to climate change but apply to data visualization and science communication in general.

\section{Conclusion} 
Visualizing climate change data in a way that makes it accessible to different audiences while still maintaining scientific accuracy is difficult. On the basis of interviews with experts and laypeople, we discussed design considerations for creating more effective climate change data visualizations, particularly in news media sources geared toward lay audiences. For our sample, we determined a need for further simplification of the shown news media (and IPCC) visualizations and/or further (textual) explanations. In comparison to the experts, \textit{lay} participants formulated longer and less abstract takeaway messages, with their sensemaking being highly influenced by design details, which eventually affected perceived attractiveness, understandability, and trustworthiness. Our findings suggest that such design details have to be carefully contemplated to fit the needs of lay audiences and we discuss such considerations along the key themes of our interviews.

\section*{Acknowledgments}
This work has been funded by the Vienna Science and Technology Fund (WWTF)[10.47379/ICT20065].

\ifCLASSOPTIONcaptionsoff
  \newpage
\fi



%

\bibliographystyle{IEEEtran}
\bibliography{IEEEabrv,bibliography}

\begin{thebibliography}{10}
\providecommand{\url}[1]{#1}
\csname url@samestyle\endcsname
\providecommand{\newblock}{\relax}
\providecommand{\bibinfo}[2]{#2}
\providecommand{\BIBentrySTDinterwordspacing}{\spaceskip=0pt\relax}
\providecommand{\BIBentryALTinterwordstretchfactor}{4}
\providecommand{\BIBentryALTinterwordspacing}{\spaceskip=\fontdimen2\font plus
\BIBentryALTinterwordstretchfactor\fontdimen3\font minus \fontdimen4\font\relax}
\providecommand{\BIBforeignlanguage}[2]{{%
\expandafter\ifx\csname l@#1\endcsname\relax
\typeout{** WARNING: IEEEtran.bst: No hyphenation pattern has been}%
\typeout{** loaded for the language `#1'. Using the pattern for}%
\typeout{** the default language instead.}%
\else
\language=\csname l@#1\endcsname
\fi
#2}}
\providecommand{\BIBdecl}{\relax}
\BIBdecl

\bibitem{bayes_research_2023}
R.~Bayes, T.~Bolsen, and J.~N. Druckman, ``A research agenda for climate change communication and public opinion: The role of scientific consensus messaging and beyond,'' \emph{Environmental Communication}, vol.~17, no.~1, pp. 16--34, 2023.

\bibitem{pew_research_center_response_2021}
{Pew Research Center}, ``In response to climate change, citizens in advanced economies are willing to alter how they live and work,'' \url{https://www.pewresearch.org/global/2021/09/14/in-response-to-climate-change-citizens-in-advanced-economies-are-willing-to-alter-how-they-live-and-work/}{ (accessed July 22, 2023)}, 2021.

\bibitem{brosch_affect_2021}
T.~Brosch, ``Affect and emotions as drivers of climate change perception and action: {A} review,'' \emph{Current Opinion in Behavioral Sciences}, vol.~42, pp. 15--21, 2021.

\bibitem{markman_why_2018}
A.~Markman, ``Why people aren’t motivated to address climate change,'' \url{https://hbr.org/2018/10/why-people-arent-motivated-to-address-climate-change}{ (accessed July 22, 2023)}, 2018.

\bibitem{joslyn_explaining_2021}
S.~Joslyn and R.~Demnitz, ``Explaining how long {CO2} stays in the atmosphere: {Does} it change attitudes toward climate change?'' \emph{Journal of Experimental Psychology: Applied}, vol.~27, no.~3, pp. 473--484, 2021.

\bibitem{head2008wicked}
B.~W. Head, ``Wicked problems in public policy,'' \emph{Public Policy}, vol.~3, no.~2, pp. 101--118, 2008.

\bibitem{johansson_evaluating_2010}
J.~Johansson, T.-S.~S. Neset, and B.-O. Linnér, ``Evaluating climate visualization: {An} information visualization approach,'' in \emph{14th {International} {Conference} on {Information} {Visualisation}, 2010, {London}, {UK}}, E.~Banissi, S.~Bertschi, R.~A. Burkhard \emph{et~al.}, Eds.\hskip 1em plus 0.5em minus 0.4em\relax IEEE Computer Society, 2010, pp. 156--161.

\bibitem{oneill_climate_2014}
S.~J. O'Neill and N.~Smith, ``Climate change and visual imagery,'' \emph{WIREs Climate Change}, vol.~5, no.~1, pp. 73--87, 2014.

\bibitem{mann2012hockey}
M.~Mann, \emph{The hockey stick and the climate wars}.\hskip 1em plus 0.5em minus 0.4em\relax Columbia University Press, 2012.

\bibitem{schneider_climate_2012}
B.~Schneider, ``Climate model simulation visualization from a visual studies perspective,'' \emph{WIREs Climate Change}, vol.~3, no.~2, pp. 185--193, 2012.

\bibitem{ipcc_summary_2021}
{IPCC}, ``Summary for {Policymakers},'' in \emph{Climate {Change} 2021: {The} {Physical} {Science} {Basis}. {Contribution} of {Working} {Group} {I} to the {Sixth} {Assessment} {Report} of the {Intergovernmental} {Panel} on {Climate} {Change}}, V.~MassonDelmotte, P.~Zhai, A.~Pirani \emph{et~al.}, Eds.\hskip 1em plus 0.5em minus 0.4em\relax Cambridge, United Kingdom and New York, NY, USA: Cambridge University Press, 2021.

\bibitem{bottinger_reflections_2020}
M.~Böttinger, H.-N. Kostis, M.~Velez-Rojas, P.~Rheingans, and A.~Ynnerman, ``\BIBforeignlanguage{en}{Reflections on visualization for broad audiences},'' in \emph{\BIBforeignlanguage{en}{Foundations of {Data} {Visualization}}}, M.~Chen, H.~Hauser, P.~Rheingans, and G.~Scheuermann, Eds.\hskip 1em plus 0.5em minus 0.4em\relax Cham: Springer International Publishing, 2020, pp. 297--305.

\bibitem{lee_reaching_2020}
B.~Lee, E.~K. Choe, P.~Isenberg, K.~Marriott, and J.~Stasko, ``Reaching broader audiences with data visualization,'' \emph{IEEE Computer Graphics and Applications}, vol.~40, no.~2, pp. 82--90, 2020.

\bibitem{peck_data_2019}
E.~M. Peck, S.~E. Ayuso, and O.~El-Etr, ``Data is personal: {Attitudes} and perceptions of data visualization in rural {Pennsylvania},'' in \emph{Proceedings of the 2019 {CHI} {Conference} on {Human} {Factors} in {Computing} {Systems}, {CHI} 2019, {Glasgow}, {Scotland}, {UK}, {May} 04-09, 2019}, S.~A. Brewster, G.~Fitzpatrick, A.~L. Cox, and V.~Kostakos, Eds.\hskip 1em plus 0.5em minus 0.4em\relax ACM, 2019.

\bibitem{burns_invisible_2022}
A.~Burns, C.~Lee, T.~On, C.~Xiong, E.~Peck, and N.~Mahyar, ``From invisible to visible: {Impacts} of metadata in communicative data visualization,'' \emph{IEEE Transactions on Visualization and Computer Graphics}, pp. 1--16, 2022.

\bibitem{park_graphoto_2018}
J.~H. Park, A.~Kaufman, and K.~Mueller, ``Graphoto: {Aesthetically} pleasing charts for casual information visualization,'' \emph{IEEE Computer Graphics and Applications}, vol.~38, no.~6, pp. 67--82, 2018.

\bibitem{kennedy_feeling_2018}
H.~Kennedy and R.~L. Hill, ``\BIBforeignlanguage{en}{The feeling of numbers: {Emotions} in everyday engagements with data and their visualisation},'' \emph{\BIBforeignlanguage{en}{Sociology}}, vol.~52, no.~4, pp. 830--848, 2018.

\bibitem{sprague_exploring_2012}
D.~Sprague and M.~Tory, ``\BIBforeignlanguage{en}{Exploring how and why people use visualizations in casual contexts: {Modeling} user goals and regulated motivations},'' \emph{\BIBforeignlanguage{en}{Information Visualization}}, vol.~11, no.~2, pp. 106--123, 2012.

\bibitem{pousman_casual_2007}
Z.~Pousman, J.~T. Stasko, and M.~Mateas, ``Casual {Information} {Visualization}: {Depictions} of {Data} in {Everyday} {Life},'' \emph{IEEE Trans. Vis. Comput. Graph.}, vol.~13, no.~6, pp. 1145--1152, 2007.

\bibitem{lee_vlat_2017}
S.~Lee, S.-H. Kim, and B.~C. Kwon, ``{VLAT}: {Development} of a visualization literacy assessment test,'' \emph{IEEE Transactions on Visualization and Computer Graphics}, vol.~23, no.~1, pp. 551--560, 2017.

\bibitem{boy_principled_2014}
J.~Boy, R.~A. Rensink, E.~Bertini, and J.-D. Fekete, ``A principled way of assessing visualization literacy,'' \emph{IEEE Transactions on Visualization and Computer Graphics}, vol.~20, no.~12, pp. 1963--1972, 2014.

\bibitem{adar_communicative_2021}
E.~Adar and E.~Lee, ``Communicative visualizations as a learning problem,'' \emph{IEEE Transactions on Visualization and Computer Graphics}, vol.~27, no.~2, pp. 946--956, 2021.

\bibitem{franconeri_science_2021}
S.~L. Franconeri, L.~M. Padilla, P.~Shah, J.~M. Zacks, and J.~Hullman, ``The science of visual data communication: {What} works,'' \emph{Psychological Science in the Public Interest}, vol.~22, no.~3, pp. 110--161, 2021.

\bibitem{lee-robbins_affective_2023}
E.~Lee-Robbins and E.~Adar, ``Affective learning objectives for communicative visualizations,'' \emph{IEEE Transactions on Visualization and Computer Graphics}, vol.~29, no.~1, pp. 1--11, 2023.

\bibitem{holder_polarizing_2023}
E.~Holder and C.~X. Bearfield, ``Polarizing political polls: How visualization design choices can shape public opinion and increase political polarization,'' \emph{IEEE Transactions on Visualization and Computer Graphics}, pp. 1--11, 2023.

\bibitem{he_enthusiastic_2023}
H.~A. He, J.~Walny, S.~Thoma, S.~Carpendale, and W.~Willett, ``Enthusiastic and grounded, avoidant and cautious: {Understanding} public receptivity to data and visualizations,'' \emph{IEEE Transactions on Visualization and Computer Graphics}, pp. 1--11, 2023.

\bibitem{morini_shock_2023}
F.~Morini, A.~Eschenbacher, J.~Hartmann, and M.~Dork, ``From shock to shift: Data visualization for constructive climate journalism,'' \emph{IEEE Transactions on Visualization and Computer Graphics}, pp. 1--11, 2023.

\bibitem{quispel_graph_2016}
A.~Quispel, A.~Maes, and J.~Schilperoord, ``Graph and chart aesthetics for experts and laymen in design: {The} role of familiarity and perceived ease of use,'' \emph{Information Visualization}, vol.~15, no.~3, pp. 238--252, 2016.

\bibitem{lee_investigation_2017}
S.~Lee, ``Investigation of visualization literacy: {A} visualization sensemaking model, a visualization literacy assessment test, and the effects of cognitive characteristics,'' Ph.D. dissertation, Purdue University, 2017.

\bibitem{landers_storytelling_2019}
K.~Giuseffi, B.~Sievert, B.~M. Wells, and F.~Westfall, ``Storytelling and sensemaking through data visualization,'' in \emph{The {Cambridge} {Handbook} of {Technology} and {Employee} {Behavior}}, R.~N. Landers, Ed.\hskip 1em plus 0.5em minus 0.4em\relax Cambridge University Press, 2019, pp. 836--846.

\bibitem{burns_designing_2022}
A.~Burns, C.~Xiong, S.~Franconeri, A.~Cairo, and N.~Mahyar, ``Designing with pictographs: {E}nvision topics without sacrificing understanding,'' \emph{IEEE Transactions on Visualization and Computer Graphics}, vol.~28, no.~12, pp. 4515--4530, 2022.

\bibitem{xiong_visual_2021}
C.~Xiong, V.~Setlur, B.~Bach, K.~Lin, E.~Koh, and S.~Franconeri, ``Visual arrangements of bar charts influence comparisons in viewer takeaways,'' \emph{IEEE Transactions on Visualization and Computer Graphics}, 2021.

\bibitem{malakis_sensemaking_2013}
S.~Malakis and T.~Kontogiannis, ``A sensemaking perspective on framing the mental picture of air traffic controllers,'' \emph{Applied Ergonomics}, vol.~44, no.~2, pp. 327--339, 2013.

\bibitem{klein_data-frame_2007}
G.~Klein, J.~Phillips, E.~Rall, and D.~Peluso, ``A data-frame theory of sensemaking,'' \emph{Expertise out of Context: Proceedings of the Sixth International Conference on Naturalistic Decision Making}, pp. 113--155, 2007.

\bibitem{russell_cost_1993}
D.~M. Russell, M.~J. Stefik, P.~Pirolli, and S.~K. Card, ``The cost structure of sensemaking,'' in \emph{Proceedings of the {INTERACT} '93 and {CHI} '93 {Conference} on {Human} {Factors} in {Computing} {Systems}}, ser. {CHI} '93.\hskip 1em plus 0.5em minus 0.4em\relax New York, NY, USA: Association for Computing Machinery, 1993, pp. 269--276.

\bibitem{burns_communicative_2022}
A.~Burns, ``Communicative information visualizations: {How} to make data more understandable by the general public,'' Ph.D. dissertation, University of Massachusetts Amherst, 2022.

\bibitem{bateman_useful_2010}
S.~Bateman, R.~L. Mandryk, C.~Gutwin, A.~Genest, D.~McDine, and C.~Brooks, ``Useful junk? the effects of visual embellishment on comprehension and memorability of charts,'' in \emph{Proceedings of the SIGCHI Conference on Human Factors in Computing Systems}.\hskip 1em plus 0.5em minus 0.4em\relax Association for Computing Machinery, 2010, p. 2573–2582.

\bibitem{gammelgaard_ballantyne_images_2016}
A.~G. Ballantyne, V.~Wibeck, and T.-S. Neset, ``Images of climate change – {A} pilot study of young people’s perceptions of {ICT}-based climate visualization,'' \emph{Climatic Change}, vol. 134, no.~1, pp. 73--85, 2016.

\bibitem{ipcc_about_nodate}
IPCC, ``{About} the {IPCC},'' \url{https://www.ipcc.ch/about/}{ (accessed July 22, 2023)}.

\bibitem{corner_principles_2018}
A.~Corner, C.~Shaw, and J.~Clarke, ``Principles for effective communication and public engagement on climate change: {A} handbook for {IPCC} authors,'' 2018, {O}xford: Climate Outreach.

\bibitem{pidcock_evaluating_2021}
R.~Pidcock, K.~Heath, L.~Messling, S.~Wang, A.~Pirani, S.~Connors, A.~Corner, C.~Shaw, and M.~Gomis, ``Evaluating effective public engagement: {L}ocal stories from a global network of {IPCC} scientists,'' \emph{Climatic Change}, vol. 168, no.~3, 2021.

\bibitem{wozniak_stakeholders_2020}
A.~Wozniak, ``Stakeholders' visual representations of climate change,'' in \emph{Research {Handbook} on {Communicating} {Climate} {Change}}.\hskip 1em plus 0.5em minus 0.4em\relax Edward Elgar Publishing, 2020, pp. 131--142.

\bibitem{harold_communication_2020}
J.~Harold, I.~Lorenzoni, T.~F. Shipley, and K.~R. Coventry, ``Communication of {IPCC} visuals: {IPCC} authors’ views and assessments of visual complexity,'' \emph{Climatic Change}, vol. 158, no.~2, pp. 255--270, 2020.

\bibitem{fischer_how_2018}
H.~Fischer, S.~Schütte, A.~Depoux, D.~Amelung, and R.~Sauerborn, ``How well do {COP22} attendees understand graphs on climate change health impacts from the {Fifth} {IPCC} {Assessment} {Report}?'' \emph{International Journal of Environmental Research and Public Health}, vol.~15, no.~5, 2018.

\bibitem{fischer_when_2020}
H.~Fischer, K.~L. van~den Broek, K.~Ramisch, and Y.~Okan, ``When {IPCC} graphs can foster or bias understanding: {E}vidence among decision-makers from governmental and non-governmental institutions,'' \emph{Environmental Research Letters}, vol.~15, no.~11, 2020.

\bibitem{mcmahon_unseen_2015}
R.~McMahon, M.~Stauffacher, and R.~Knutti, ``The unseen uncertainties in climate change: {Reviewing} comprehension of an {IPCC} scenario graph,'' \emph{Climatic Change}, vol. 133, no.~2, pp. 141--154, 2015.

\bibitem{harold_enhancing_2017}
J.~Harold, I.~Lorenzoni, K.~Coventry, and A.~Minns, ``\BIBforeignlanguage{en}{Enhancing the accessibility of climate change data visuals: {Recommendations} to the {IPCC} and guidance for researchers},'' \url{https://tyndall.ac.uk/wp-content/uploads/2021/09/Data_Visuals_Guidance_Full_Report_0.pdf}{ (accessed July 22, 2023)}, 2017.

\bibitem{gaulkin_why_2021}
T.~Gaulkin, ``{Why} the bad news in the {IPCC} report is good news for visual learners,'' \url{https://thebulletin.org/2021/08/why-the-bad-news-in-the-ipcc-report-is-good-news-for-visual-learners/} { (accessed July 22, 2023)}, 2021.

\bibitem{harold_cognitive_2016}
J.~Harold, I.~Lorenzoni, T.~F. Shipley, and K.~R. Coventry, ``Cognitive and psychological science insights to improve climate change data visualization,'' \emph{Nature Climate Change}, vol.~6, no.~12, pp. 1080--1089, 2016.

\bibitem{oneill_more_2020}
S.~O’Neill, ``More than meets the eye: {A} longitudinal analysis of climate change imagery in the print media,'' \emph{Climatic Change}, vol. 163, no.~1, pp. 9--26, 2020.

\bibitem{hopke_visualizing_2018}
J.~E. Hopke and L.~E. Hestres, ``Visualizing the {Paris} {Climate} {Talks} on {Twitter}: {Media} and climate stakeholder visual social media during {COP21},'' \emph{Social Media + Society}, vol.~4, no.~3, 2018.

\bibitem{wessler_global_2016}
H.~Wessler, A.~Wozniak, L.~Hofer, and J.~Lück, ``Global multimodal news frames on climate change: {A} comparison of five democracies around the world,'' \emph{The International Journal of Press/Politics}, vol.~21, no.~4, pp. 423--445, 2016.

\bibitem{wang_public_2018}
S.~Wang, A.~Corner, D.~Chapman, and E.~Markowitz, ``Public engagement with climate imagery in a changing digital landscape,'' \emph{Wiley Interdisciplinary Reviews: Climate Change}, vol.~9, no.~2, 2018.

\bibitem{feldman_is_2018}
L.~Feldman and P.~S. Hart, ``Is there any hope? {How} climate change news imagery and text influence audience emotions and support for climate mitigation policies,'' \emph{Risk Analysis}, vol.~38, no.~3, pp. 585--602, 2018.

\bibitem{metag_perceptions_2016}
J.~Metag, M.~S. Schäfer, T.~Füchslin, T.~Barsuhn, and K.~Kleinen-von Königslöw, ``Perceptions of climate change imagery: {Evoked} salience and self-efficacy in {Germany}, {Switzerland}, and {Austria},'' \emph{Science Communication}, vol.~38, no.~2, pp. 197--227, 2016.

\bibitem{windhager_inconvenient_2019}
F.~Windhager, G.~Schreder, and E.~Mayr, ``On inconvenient images: {Exploring} the design space of engaging climate change visualizations for public audiences,'' in \emph{7th {Workshop} on {Visualisation} in {Environmental} {Sciences}, {EnvirVis}@{EuroVis} 2019, {Porto}, {Portugal}, {June} 3, 2019}, R.~Bujack, K.~Feige, K.~Rink \emph{et~al.}, Eds.\hskip 1em plus 0.5em minus 0.4em\relax Eurographics Association, 2019, pp. 1--8.

\bibitem{nocke_visualization_2008}
T.~Nocke, T.~Sterzel, M.~Böttinger, and M.~Wrobel, ``Visualization of climate and climate change data: {An} overview,'' \emph{in Ehlers et al. (Eds.) Digital Earth Summit on Geoinformatics 2008: Tools for Global Change Research (ISDE'08), Wichmann, Heidelberg, pp. 226-232}, 2008.

\bibitem{dudman_ipcc_2021}
K.~Dudman and S.~de~Wit, ``An {IPCC} that listens: {Introducing} reciprocity to climate change communication,'' \emph{Climatic Change}, vol. 168, no.~1, 2021.

\bibitem{BBC}
M.~McGrath, ``{Climate} change: {Five things} we have learned from the {IPCC} report,'' \url{https://www.bbc.com/news/science-environment-58138714}{ (accessed July 22, 2023)}, 2021, {BBC} News at bbc.co.uk/news.

\bibitem{the_guardian_as_2021}
Guardian, ``As a verdict on the climate crimes of humanity, the new {Intergovernmental} {Panel} on {Climate} {Change} report could not be clearer: {W}e're guilty as hell.'' \url{https://www.instagram.com/p/CSXHuKIqCs1/}{ (accessed July 22, 2023)}, 2021, {S}lide 4.

\bibitem{schuster2023the}
R.~Schuster, L.~Koesten, K.~Gregory, and T.~Möller, ```{T}he main message is that sustainability would help' -- {R}eflections on takeaway messages of climate change data visualizations,'' {\url{ https://doi.org/10.48550/arXiv.2305.04030} (accessed Jan 3, 2024)}, 2023.

\bibitem{robson2017real}
C.~Robson and K.~McCartan, ``Real world research, 4th edition,'' 2017.

\bibitem{kekeya_analysing_2020}
J.~Kekeya, ``Analysing qualitative data using an iterative process,'' \emph{Contemporary PNG Studies}, vol.~24, pp. 86--94, 2020.

\bibitem{ladonna_beyond_2021}
K.~A. LaDonna, A.~R. Artino, Jr, and D.~F. Balmer, ``Beyond the guise of saturation: {Rigor} and qualitative interview data,'' \emph{Journal of Graduate Medical Education}, vol.~13, no.~5, pp. 607--611, 2021.

\bibitem{IPCC-slides}
IPCC, ``{AR}6 {WGI SPM} {B}asic slide deck with figures,'' \url{https://www.ipcc.ch/report/ar6/wg1/downloads/outreach/IPCC_AR6_WGI_SPM_Basic_Slide_Deck_Figures.pdf} { (accessed Mar. 29, 2021)}, 2021.

\bibitem{noauthor_quarks_2022}
@quarks {\textbar}~{Instagram}, ``So hat sich die globale {Temperatur} der {Erde} entwickelt,'' \url{https://www.instagram.com/p/CbnFrMIK9fl/}{ (accessed July 22, 2023)}, 2022, {Author}: Lena Bültena, Editor: Andreas Sträter, Graphic Artist: Elica Petrova-Wallek.

\bibitem{noauthor_tagesschau_nodate}
@tagesschau {\textbar}~{Instagram}, \url{https://www.instagram.com/tagesschau/}{ (accessed July 22, 2023)}.

\bibitem{williams_non-designers_2017}
R.~Williams, \emph{The non-designer's presentation book: {Principles} for effective presentation design}.\hskip 1em plus 0.5em minus 0.4em\relax Peachpit Press, 2017.

\bibitem{peck_personal_2019}
E.~M. Peck, S.~E. Ayuso, and O.~El-Etr, ``Data is personal: Attitudes and perceptions of data visualization in rural pennsylvania,'' in \emph{Proceedings of the 2019 CHI Conference on Human Factors in Computing Systems}, ser. CHI '19.\hskip 1em plus 0.5em minus 0.4em\relax Association for Computing Machinery, 2019, p. 1–12.

\bibitem{hullman_visualization_2011}
J.~Hullman and N.~Diakopoulos, ``Visualization rhetoric: Framing effects in narrative visualization,'' \emph{IEEE Transactions on Visualization and Computer Graphics}, vol.~17, no.~12, pp. 2231--2240, 2011.

\bibitem{borkin_memorability_2016}
M.~A. Borkin, Z.~Bylinskii, N.~W. Kim, C.~M. Bainbridge, C.~S. Yeh, D.~Borkin, H.~Pfister, and A.~Oliva, ``Beyond memorability: Visualization recognition and recall,'' \emph{IEEE Transactions on Visualization and Computer Graphics}, vol.~22, no.~1, pp. 519--528, 2016.

\bibitem{kong_frames_2018}
H.-K. Kong, Z.~Liu, and K.~Karahalios, ``Frames and slants in titles of visualizations on controversial topics,'' in \emph{Proceedings of the 2018 CHI Conference on Human Factors in Computing Systems}, ser. CHI '18.\hskip 1em plus 0.5em minus 0.4em\relax Association for Computing Machinery, 2018, p. 1–12.

\bibitem{kause_visualizations_2020}
A.~Kause, W.~Bruine~de Bruin, F.~Fung \emph{et~al.}, ``Visualizations of projected rainfall change in the {U}nited {K}ingdom: {An} interview study about user perceptions,'' \emph{Sustainability}, vol.~12, 2020.

\bibitem{mcmahon_scientific_2016}
R.~McMahon, M.~Stauffacher, and R.~Knutti, ``The scientific veneer of {IPCC} visuals,'' \emph{Climatic Change}, vol. 138, no.~3, pp. 369--381, 2016.

\bibitem{ferreira2021Climate}
M.~Ferreira, M.~Coelho, V.~Nisi, and N.~Jardim~Nunes, ``Climate change communication in {HCI}: {A} visual analysis of the past decade,'' in \emph{Creativity and {Cognition}}.\hskip 1em plus 0.5em minus 0.4em\relax ACM, 2021, pp. 1--16.

\bibitem{burns2020how}
A.~Burns, C.~Xiong, S.~Franconeri, A.~Cairo, and N.~Mahyar, ``How to evaluate data visualizations across different levels of understanding,'' in \emph{2020 IEEE Workshop on Evaluation and Beyond - Methodological Approaches to Visualization (BELIV)}, 2020.

\bibitem{rothermich_influence_2021}
K.~Rothermich, E.~K. Johnson, R.~M. Griffith, and M.~M. Beingolea, ``The influence of personality traits on attitudes towards climate change – {An} exploratory study,'' \emph{Personality and Individual Differences}, vol. 168, 2021.

\bibitem{Deng2016}
S.~Deng and V.~Sloutsky, ``Selective attention, diffused attention, and the development of categorization,'' \emph{Cognitive Psychology}, vol.~91, 2016.

\bibitem{lee_how_2016}
S.~Lee, S.-H. Kim, Y.-H. Hung, H.~Lam, Y.-A. Kang, and J.~S. Yi, ``How do people make sense of unfamiliar visualizations?: {A} grounded model of novice's information visualization sensemaking,'' \emph{IEEE Trans. Vis. Comput. Graph.}, vol.~22, no.~1, pp. 499--508, 2016.

\bibitem{fish_storytelling_2020}
C.~Fish, ``Storytelling for making cartographic design decisions for climate change communication in the {United} {States},'' \emph{Cartographica: The International Journal for Geographic Information and Geovisualization}, vol.~55, no.~2, pp. 69--84, 2020, publisher: University of Toronto Press.

\bibitem{hinds_bothered_2001}
P.~J. Hinds, M.~Patterson, and J.~Pfeffer, ``\BIBforeignlanguage{eng}{Bothered by abstraction: {The} effect of expertise on knowledge transfer and subsequent novice performance},'' \emph{\BIBforeignlanguage{eng}{The Journal of Applied Psychology}}, vol.~86, no.~6, pp. 1232--1243, 2001.

\bibitem{lundgard_accessible_2021}
A.~Lundgard and A.~Satyanarayan, ``Accessible visualization via natural language descriptions: {A} four-level model of semantic content,'' \emph{IEEE Transactions on Visualization and Computer Graphics}, 2021.

\bibitem{morelli_co-designing_2021}
A.~Morelli, T.~G. Johansen, R.~Pidcock, J.~Harold, A.~Pirani, M.~Gomis, I.~Lorenzoni, E.~Haughey, and K.~Coventry, ``Co-designing engaging and accessible data visualisations: {A} case study of the {IPCC} reports,'' \emph{Climatic Change}, vol. 168, no.~3, p.~26, 2021.

\bibitem{fischhoff_sciences_2013}
B.~Fischhoff, ``The sciences of science communication,'' \emph{Proceedings of the National Academy of Sciences}, vol. 110, 2013.

\bibitem{samuelson_status_1988}
W.~Samuelson and R.~Zeckhauser, ``Status quo bias in decision making,'' \emph{Journal of Risk and Uncertainty}, vol.~1, no.~1, pp. 7--59, 1988.

\bibitem{patt_action_2000}
A.~Patt and R.~Zeckhauser, ``Action bias and environmental decisions,'' \emph{Journal of Risk and Uncertainty}, vol.~21, no.~1, pp. 45--72, 2000.

\bibitem{boehm_state_2021}
S.~Boehm, K.~Lebling, K.~Levin \emph{et~al.}, ``State of climate action 2021: {S}ystems transformations required to limit global warming to 1.5°{C},'' 2021.

\bibitem{fish_cartographic_202}
C.~S. Fish, ``Cartographic content analysis of compelling climate change communication,'' \emph{Cartography and Geographic Information Science}, vol.~47, no.~6, pp. 492--507, 2020.

\bibitem{burns_novices_2023}
A.~Burns, C.~Lee, R.~Chawla, E.~Peck, and N.~Mahyar, ``Who do we mean when we talk about visualization novices?'' in \emph{Proceedings of the 2023 CHI Conference on Human Factors in Computing Systems}.\hskip 1em plus 0.5em minus 0.4em\relax Association for Computing Machinery, 2023.

\end{thebibliography}

\end{document}